\documentstyle[manuscript,aps,epsfig]{revtex}
\catcode`\@=11
\newcommand{\fcaption}[1]{\vspace{1ex}
        \refstepcounter{figure}
        \setbox\@tempboxa = \hbox{\footnotesize {\bf ig.~\thefigure.} #1}
        \ifdim \wd\@tempboxa > 8cm
           {\begin{center}
        \parbox{8cm}{\footnotesize\baselineskip=8pt {\bf Fig.~\thefigure.} #1}

            \end{center}}
        \else
             {\begin{center}
             {\footnotesize {\bf Fig.~\thefigure.} #1}
              \end{center}}
        \fi}
\newcommand{\be}{\begin{equation}}
\newcommand{\ee}{\end{equation}}
\newcommand{\bea}{\begin{eqnarray}}
\newcommand{\eea}{\end{eqnarray}}

\begin{document}

\centerline{\large \bf Zero-Variance Zero-Bias Principle for Observables in 
quantum Monte Carlo:}
\centerline{\large \bf Application to Forces}

\vspace{2cm}

\begin{center}
Roland Assaraf and Michel Caffarel
\end{center}
\vspace{1cm}
\noindent
{CNRS-Laboratoire de Chimie Th\'eorique Tour 22-23, Case 137\\
Universit\'{e} Pierre et Marie Curie, 4 place Jussieu 75252 Paris Cedex 05 France\\
e-mail: ra@lct.jussieu.fr,mc@lct.jussieu.fr}

\begin{center}
{\bf Abstract}
\end{center}
A simple and stable method for computing accurate expectation values 
of observable with Variational Monte Carlo (VMC) or Diffusion Monte 
Carlo (DMC) algorithms is presented. The basic idea consists in
replacing the usual ``bare'' estimator associated with the observable 
by an improved or ``renormalized'' estimator. Using this estimator
more accurate averages are obtained:
Not only the statistical fluctuations are reduced but also 
the systematic error (bias) associated with the approximate VMC 
or (fixed-node) DMC probability densities. It is shown that improved estimators 
obey a Zero-Variance Zero-Bias (ZVZB) property similar 
to the usual Zero-Variance Zero-Bias property of the energy
with the local energy as improved estimator.
Using this property improved estimators can be optimized 
and the resulting accuracy on expectation values may reach
the remarkable accuracy obtained for total energies.
As an important example, we present the application of our formalism
to the computation of forces in molecular systems.
Calculations of the entire force curve of the H$_2$,LiH, and Li$_2$ molecules
are presented. Spectroscopic constants $R_e$ (equilibrium distance) 
and $\omega_e$ (harmonic frequency) are also computed. 
The equilibrium distances are obtained with a relative error smaller than 1$\%$, while 
the harmonic frequencies are computed with an error of about 10$\%$. 

\newpage

\section{Introduction}

 Over the recent years quantum Monte Carlo (QMC) methods have become 
more and more successful in computing ground-state total energies of 
molecular systems. For systems with large number of electrons the 
accuracy obtained by QMC is very good. As illustrated by a number of 
recent calculations,
\cite{wil1,wil2,wil3,mitas1,fil1,needs1,mitas2,lester1,wil4,fil2,needs2,mitas3,mitas4,anderson0} 
the quality of the results is comparable 
and, in most cases, superior to that obtained with more traditional 
techniques (DFT, MCSCF or coupled cluster methods). Unfortunately, 
for properties other than energy the situation 
is much less favorable and accurate results are difficult 
to obtain. To understand this point let us first define what 
we mean here by accuracy. In standard quantum Monte Carlo schemes
there exist essentially two types of error:

i) the usual statistical error resulting from the necessarily finite
simulation time. This error present in any Monte Carlo scheme behaves 
as $ \sim 1/\sqrt{N}$ where $N$ is the number of Monte Carlo steps.

ii) the systematic error (or ``bias'') associated with some particular 
choice of the trial wave function. 
In a Variational Monte Carlo (VMC)
scheme it is the systematic error resulting from the approximate trial probability 
density. In a fixed-node Diffusion Monte Carlo (DMC)
it is either the fixed-node error of energy calculations or the systematic error 
associated with the mixed DMC probability density 
for a more general observable.
Other types of systematic errors may also exist, e.g. the short-time 
error, \cite{lester2} however, such errors can be easily controlled and, therefore,
will not be considered here.

Now, to enlighten the major differences between energy and observable 
computations let us compute the expressions of these two errors.
We shall do that within the framework of the Variational Monte Carlo
method where, as we shall see later, all the main aspects of this work 
are already present.

In a variational Monte Carlo simulation the variational energy
\begin{equation}
E_v \equiv \frac { \langle \psi_T | H | \psi_T \rangle} 
                 { \langle \psi_T |     \psi_T \rangle},
\label{variationalenergy}
\end{equation}
where $\psi_T$ is the approximate trial wave function used, is re-expressed 
as the statistical average of the local energy defined as
\begin{equation} 
E_L = \frac {H\psi_T}{\psi_T} 
\label{localenergy}
\end{equation} 
over the probability density associated with $\psi_T^2$, namely
\begin{equation}
E_v = \langle E_L \rangle_{\psi_T^2}.
\label{averagenergy}
\end{equation}
An accurate calculation of the energy requires the two following conditions.

(i) First, the systematic (or variational) error defined as 
\begin{equation}
\Delta_E  \equiv E_v - E_0  \ge 0 ,
\label{deltadef}
\end{equation}
where $E_0$ is the exact energy, must be as small as possible.

(ii) Second, the variance of the local energy (which is directly related to the 
magnitude of the statistical error)
\begin{equation}
\sigma^2(E_L)= \langle (E_L-E_v)^2 \rangle_{\psi_T^2},
\label{vardef}
\end{equation}
must also be as small as possible.

To estimate both quantities we express them in terms of the 
trial wave function error, $\delta \psi = \psi_T-\psi_0$, where $\psi_0$
is the exact wave function.
Regarding the systematic error it is easy to check that 
\begin{equation} 
\Delta_E  = \frac {\langle \psi_T-\psi_0| H - E_0 | \psi_T-\psi_0 \rangle}
                  {\langle \psi_T|\psi_T \rangle }.
\label{delta}
\end{equation} 
In other words, $\Delta_E$ is of order two in the wave function error
\begin{equation}    
\Delta_E  \sim  O[{\delta \psi}^2].
\label{deltaorder}
\end{equation}
Now, regarding the variance, it is convenient to write the 
following equality
\begin{equation} 
E_L - E_v =  \frac { (H - E_0) (\psi_T-\psi_0) } {\psi_T} - \Delta_E,
\label{eqint1}
\end{equation}
from which it is directly seen that $\sigma^2(E_L)$ is also of order two 
\begin{equation}
\sigma^2(E_L) \sim  O[{\delta \psi}^2].
\label{varorder}
\end{equation} 

Equations (\ref{deltaorder}) and (\ref{varorder}) are at the origin of 
the high-quality calculations of the energy. 
They show that accurate energy calculations are directly related to 
good trial wave functions: The more accurate the
trial wave function is, the smaller the statistical and systematic 
errors are. In the limit of an exact trial wave function, both errors 
vanish and the energy estimator reduces to the exact energy. 
This most fundamental property is referred to in the literature as 
the ``Zero-Variance property''. Note that a much more preferable 
and accurate denomination should be ``Zero-Variance-Zero-Bias property" 
to emphasize on the existence of the {\it two} types of error. Of course, in the 
case of the energy this distinction is not necessary since, as just
seen, the two errors are not independent and vanish
{\it simultaneously} with the exact wave function.
However, as we shall see below, this peculiar aspect 
will be no longer true for other properties.

Let us now turn our attention to the computation of a general observable.
Defining the expectation value of some arbitrary observable $O$ 
as follows
\begin{equation}
O_v \equiv \frac {\langle \psi_T | O | \psi_T \rangle } {\langle \psi_T
  | \psi_T \rangle},
\label{variationalobs}
\end{equation}
its Monte Carlo expression is given by
\begin{equation}
O_v = \langle O \rangle_{\psi_T^2}.
\label{averageobs}
\end{equation}
It is easy to verify that the systematic error behaves as
\begin{equation} 
\Delta_O \equiv \langle O \rangle_{\psi_T^2} - 
\langle O \rangle_{\psi_0^2} \sim  O[\delta \psi],
\label{deltaobs}
\end{equation}
while the variance is given by
\begin{equation}
\sigma^2(O) \sim  O[1].
\label{varorderobs}
\end{equation}
Compared to the energy case we have two striking differences.
First, the systematic error in the averages is much larger.
This is a direct consequence of Eq.(\ref{deltaobs}):
the estimator of a general observable has only a {\it linear} zero-bias property 
instead of a quadratic one like in the energy case. Even worse, because
trial wave functions are optimized to lower the
systematic error in the energy (and/or its fluctuations) and not
the error in the observable, the prefactor associated with the linear error 
contribution, Eq.(\ref{deltaobs}), is usually much 
larger than in the energy case, Eq.(\ref{deltaorder}). 
In practice, this important systematic error makes in general the quality of 
the expectation value, Eq.(\ref{averageobs}), very poor.
The second important difference is that there is no zero-variance property at all 
for observables when Eq.(\ref{averageobs}) is used. Indeed,
even when the exact wave function is used as trial wave function
we are still left with some finite (and eventually large) statistical fluctuations,
Eq.(\ref{varorderobs}). Thus, statistical fluctuations are in general very large 
for properties.
A simple and popular strategy to reduce the important systematic 
error on properties is to mix Variational Monte Carlo (VMC) and fixed-node Diffusion 
Monte Carlo (DMC) calculations to build up a so-called ``hybrid'' or ``second-order'' estimator,
${\langle O \rangle}_{hybrid} \equiv 2{\langle O \rangle }_{DMC}
- {\langle O \rangle }_{VMC}$, whose error
is reduced.\cite{ceperley1} An elementary calculation shows that
the error is now of order $O[{(\psi_T-\psi_0)}^2]$, plus a linear
contribution $O(\psi_0^{FN}-\psi_0)$ due to the approximate nodes
of the trial wave function. However, once again such a solution is not, in practice,
as satisfactory as it appears at first glance because of the large 
prefactor associated with the second-order contribution and, also, of 
the non-negligible linear error due to the nodes.
A second possible strategy to cope with the systematic error is to perform
an ``exact'' QMC calculation based on one of the variants of the so-called 
``Forward Walking'' scheme.\cite{liu1,caffaclav,casul1} Unfortunately, such schemes are known to be
intrinsically unstable and, therefore, very time consuming. In practice, 
the possibility of getting or not a satisfactory answer depends very much on 
the accuracy required and on the type of observable considered. 
Therefore, Forward Walking is not considered as a general practical 
solution to the problem.

In this work, we propose to follow a quite different route. Our purpose is 
to show that it is possible to use much more efficient estimators for 
properties than the usual bare expression, Eq.(\ref{averageobs}). 
More precisely, it is shown how to construct in a simple and systematic way
new estimators having the same remarkable quadratic zero-variance zero-bias 
property as the energy case.
Very recently, we have made a first step in that direction by
showing how to generalize the zero-variance part of this property.\cite{zvp1,zvp2}
In short, the basic idea consists in constructing a ``renormalized'' or improved 
observable having the same average as the original one but a lower variance. 
To build the renormalized observable, an auxiliary wave function 
is introduced. This function plays a role analogous to the one played 
by the trial wave function in the case of the energy: The closer the auxiliary 
function is of the exact solution of some zero-variance equation 
(the Schroedinger equation in the case of the energy), the smaller 
the statistical fluctuations of the renormalized observable are. 
Our approach has been illustrated on some simple academic examples\cite{zvp1}
and also for the much more difficult case of the computation of forces
for some diatomic molecules.\cite{zvp2} Numerical results on these 
examples are very satisfactory. When suitably chosen auxiliary functions 
are used, statistical errors are indeed greatly reduced.

Here, we present the full generalization of the preceding idea: 
it is shown how to construct improved observables 
minimizing {\it both} systematic and statistical
errors with a quadratic behavior similar to that obtained for the energy. 
As a consequence, any observable is expected to be calculated, at least 
in principle, with the remarkable accuracy achieved by QMC for total energies.
The basic idea behind our approach is quite simple: it consists in 
making use of the relation between energy and observable calculations 
as expressed by the Hellmann-Feynman (HF) theorem. As well-known this 
theorem expresses any quantum average as a total energy derivative with 
respect to the magnitude of the external potential defined by the observable.
It is shown how the zero-variance zero-bias principle valid for 
each value of the energy (as a function of the external potential)
can be extended to the derivative and, therefore, to the observable.
Note that in the context of QMC simulations, the idea of using the 
HF theorem to compute observables, using either a finite difference scheme or 
the analytic derivative, is not new and has been applied by several 
groups\cite{wells1,traynor1,umrigar1,sun1,vrbik1,vrbik2,belohorec1,fil3,baer}.
In general, the results are good for very small systems but rapidly disappointing 
for larger systems. Indeed, only when a clear physical 
insight into the origin of the fluctuations of the infinitesimal difference 
of energy (the derivative) is available it is possible to propose an 
efficient solution to the problem. 
A very nice example of such a possibility 
is presented in the recent work by Filippi and Umrigar.
\cite{umrigarwrapped},\cite{fil3}. By using a finite representation of 
the energy derivative and by introducing a special coordinate transformation 
allowing the electrons close to a given nucleus to move almost rigidly 
with that nucleus, they have shown how to correlate efficiently
the calculation of the electronic energies associated with two slighlty 
different nuclear configurations of a diatomic molecule.
As a result they have been able to get accurate estimates of the 
energy derivatives (forces) for some diatomic molecules. 
Here, we show how the correlated sampling 
method of Filippi and Umrigar can be re-expressed in our framework. 
In addition, by generalizing their idea it is shown how coordinate 
transformations can be used to define a new class of improved estimators. 
 
While finishing this work, we came aware of a paper just published by Casalegno, 
Mella and Rappe.\cite{cmr} The idea underlying their work has some 
close relations with what is presented here. In short, they propose, as we do here, 
to compute forces using a Hellmann-Feynman-type formalism. 
Their expression to calculate forces is obtained by making the derivative 
of the VMC (or DMC) energy average with respect to nuclear positions.
To reduce the systematic error these authors propose to employ trial wave 
functions which have been very carefully optimized via energy minimization 
(let us recall that the HF theorem is valid when {\it fully} optimized wave functions are used). 
To decrease the very large statistical fluctuations associated with the infinite 
variance, the improved estimator introduced in our previous work 
on forces\cite{zvp2} is used. As we shall see below, the approach proposed by 
Casalegno and collaborators can be viewed as a special case of the general method presented here, except 
that their estimator does not obey a Zero-Variance Zero-Bias property. As we shall see below, this latter aspect 
has some important practical consequences when a high level of accuracy on forces is needed.

The organization of the paper is as follows. In Section II we present 
the Hellmann-Feynman theorem and the construction of improved estimators 
for variational Monte Carlo calculations.
It is also shown how the idea of Filippi and
Umrigar consisting in introducing a special coordinate transformation can be 
used to build up some more general and more efficient improved estimators.
In Section III we discuss the generalization of the formulae to the 
case of Diffusion Monte Carlo calculations.
In Section IV we present the application of the formalism
to the computation of the entire force curve 
for the H$_2$,LiH, and Li$_2$ molecules.
Calculations of the spectroscopic constants, $R_e$ 
and $\omega_e$, are also reported.
Finally, in the last section we summarize our results and present 
some concluding remarks.

\section{Improved estimators for observables} 

In order to make the connection between energy and observable computations 
we shall make use of the Hellmann-Feynman (HF) theorem which expresses
the expectation value of an observable as an energy derivative 
\begin{equation}
\frac{ \langle \psi_0|O|\psi_0 \rangle}
{\langle \psi_0|\psi_0 \rangle} 
= \frac{d E_0(\lambda)}{d\lambda}\big|_{\lambda=0},
\label{eq:exact}
\end{equation}
where $E_0(\lambda)$ is the exact ground-state energy of 
the ``perturbed'' Hamiltonian defined as
\begin{equation}
H(\lambda) \equiv H + \lambda O.
\end{equation}
By choosing various approximate expressions for 
the exact energy in Eq.(\ref{eq:exact}), it is possible to derive various
approximate estimates for the average. In the next sections we present
two choices which turn out to be particularly efficient 
in practical applications.

\subsection{Improved estimator built from the variational approximation of 
the energy}

A most natural choice consists in replacing the exact energy 
of the HF theorem by a high-quality variational approximation. 
To do that, we introduce some $\lambda$-dependent approximate trial 
wave function, $\psi_T(\lambda)$ to describe the ground-state of 
$H(\lambda)$ 
[Note that, for the sake of clarity and simplicity, we shall denote 
in what follows ${\psi_T}(0)$, $H(0)$, and $E_0(0)$ as ${\psi_T}$, $H$, 
and $E_0$, respectively]. 

The exact average of the observable can be decomposed as
\begin{equation}
\frac{ \langle \psi_0|O|\psi_0 \rangle } 
{\langle \psi_0|\psi_0 \rangle }
=\frac{d {E_v}(\lambda)} {d\lambda}  \big|_{\lambda=0} 
+ \epsilon(\delta \psi,\delta \psi^\prime)
\label{approx}
\end{equation}
where ${E_v}(\lambda)$ is the variational energy associated with 
$\psi_T(\lambda)$
\begin{equation}
{E_v}(\lambda) \equiv \langle {E_L}(\lambda) \rangle_{{\psi_T}^2(\lambda)} = 
\langle \frac{H(\lambda) {\psi_T}(\lambda)}{{\psi_T}(\lambda)} \rangle_{{\psi_T}^2(\lambda)}
\end{equation}
and $\epsilon$ some correction depending on $\delta\psi=\psi_0-\psi_T$ and 
its derivative, and vanishing when the exact wave function is used as 
trial wave function.

Now, the important point is that the derivative of the variational energy, 
$\frac{d{E_v}(\lambda)} {d\lambda} \big|_{\lambda=0}$,
is expected to be a better estimate of the exact average than the ordinary 
average of the bare estimator, Eq.(\ref{averageobs}), when properly chosen 
$\lambda$-dependent trial wave functions are used. This is true since
the standard estimator, Eq.(\ref{averageobs}), can be re-expressed as 
a particular case of the derivative of the variational energy for 
a $\lambda$-{\it independent} trial wave function, 
a choice which is clearly not optimal.
Before justifying more quantitatively this statement, let 
us rewrite the derivative as an ordinary 
average over the density ${\psi_T}^2$. This can be easily done, it gives
\begin{equation}
\frac{d{E_v}(\lambda)}{d\lambda} \big|_{\lambda=0} 
= \langle \tilde{O} \rangle_{{\psi_T}^2}
\label{evprime}
\end{equation}
where $\tilde{O}$ is a new modified local operator written as
\begin{equation}
\tilde{O} \equiv  O + \frac{(H-{E_L}){\psi_T}^\prime}{{\psi_T}}
+2 ({E_L}-{E_v}) \frac{{{\psi_T}}^\prime}
{{\psi_T}}
\label{tildeobias}
\end{equation}
In this latter formula, and in the formulae to follow, we shall use 
the following simplified notation  
\begin{equation}
f^\prime \equiv \frac{d f(\lambda)}{d\lambda} 
\big|_{\lambda =0}
\end{equation}
where $f(\lambda)$ is some arbitrary function of $\lambda$.

Now, we have to justify the first important result,
that the new estimator $\tilde{O}$ is a 
better estimator for the exact average than the bare observable $O$. 
For that purpose, we compute the systematic error in the corresponding 
average and the variance of the new operator.
Regarding the systematic error we can write
\begin{equation}
\Delta_{\tilde O} \equiv
\langle \tilde O \rangle_{{{\psi_T}}^2} - \langle O \rangle_{\psi_0^2}
= \frac{d[{E_v}(\lambda)-E_0(\lambda)]}{d\lambda} \big|_{\lambda=0}.
\end{equation}

Let us denote ${\psi_0}(\lambda)$ the exact groundstate of $H(\lambda)$ 
[with $\psi_0(0)=\psi_0$]. Using the equality
$$
E_v(\lambda)-E_0(\lambda)=
$$
\begin{equation}
\frac {\langle {\psi_T}(\lambda) - {\psi_0}(\lambda)| H(\lambda) - E_0(\lambda) 
 | {\psi_T}(\lambda) - {\psi_0}(\lambda)  \rangle}
                  {\langle {\psi_T}(\lambda)|{\psi_T}(\lambda) \rangle }
\end{equation}
and choosing the following convention of normalization 
\begin{equation}
\langle {\psi_T}(\lambda)|{\psi_T}(\lambda) \rangle= 1,
\nonumber
\end{equation}
the derivative can be easily computed, we get
\begin{eqnarray}
 \Delta_{\tilde O} & = & 
 \langle {\psi_T}
-\psi_0 | O - \langle O \rangle_{\psi_0^2} |  {\psi_T} -\psi_0 \rangle  
\nonumber \\ 
& + &
 2 \langle {\psi_T}-\psi_0 | H -E_0 | {{\psi_T}}^\prime-{\psi_0}^\prime \rangle 
\label{biasderiv}
\end{eqnarray}
As it can be seen, the systematic error is now of order two in the errors ${\psi_T}-\psi_0$ and
${{\psi_T}}^\prime-{\psi_0}^\prime$
\begin{equation}
\Delta_{\tilde O} \sim O[({\psi_T}-\psi_0)
({\psi_T}^\prime-{\psi_0}^\prime)].
\label{eqint3}
\end{equation}

Now, let us compute the variance defined as
\begin{equation}
\sigma^2(\tilde{O}) = \langle 
(\tilde{O} - \langle \tilde{O} \rangle_{{\psi_T}^2})^2 \rangle_{{\psi_T}^2}.
\label{varotilde}
\end{equation}
Using Eqs. (\ref{evprime}),(\ref{tildeobias}) and the fact that 
$E_L^\prime= O +  \frac{(H-{E_L}){\psi_T}^\prime}{{\psi_T}}$
we can express the difference $\tilde{O} - 
\langle \tilde{O}\rangle_{{\psi_T}^2}$ as follows
\begin{equation}
\tilde{O}- \langle{\tilde{O}} 
\rangle_{{{\psi_T}}^2}  =  {E_L}^\prime -{E_v}^\prime
+2 ({E_L} -{E_v}) \frac{{\psi_T}^\prime}
{{\psi_T}}. 
\label{eqint2}
\end{equation} 
For the sake of clarity, let us distinguish two different contributions 
in the difference. The first contribution is given by
\begin{equation}
{E_L}^\prime -{E_v}^\prime = \frac{d[{E_L}(\lambda) -{E_v}(\lambda)]}
{d\lambda}\mid_{\lambda=0}.
\end{equation}
Using expression (\ref{eqint1}) for $[{E_L}(\lambda)-{E_v}(\lambda)]$ 
and performing the derivative one obtains
$$
{E_L}^\prime -{E_v}^\prime = \frac{(O-\langle O \rangle_{\psi_0^2})
 ({\psi_T}-\psi_0)}{{\psi_T}}  
$$
$$
 +
\frac{(H-E_0)({\psi_T}^\prime-\psi_0^\prime)}{{\psi_T}} 
 - 
\frac{(H-E_0)({\psi_T}-\psi_0)}{{\psi_T}} \frac{ {\psi_T}^\prime }{\psi_T}
$$

\begin{equation}
 + 
\langle O \rangle_{\psi_0^2}- \langle \tilde{O} \rangle_{\psi_T^2}
\end{equation}
This latter expression is clearly of order one 
in ${{\psi_T}}-{\psi_0}$ and its derivative,${\psi_T}^\prime-\psi_0^\prime $.
The second contribution in the R.H.S. of Eq.(\ref{eqint2}) is 
proportional to ${E_L} -{E_v}$. We have already seen that 
it is of order one in ${\psi_T}-{\psi_0}$, Eqs.(\ref{deltaorder}),
(\ref{eqint1}). Finally, $\tilde{O}-\langle\tilde{O}\rangle_{{\psi_T}^2}$
is found to be of order one in 
${\psi_T}-\psi_0$ and ${\psi_T}^\prime-\psi_0^\prime$.
The variance, Eq.(\ref{varotilde}), is therefore of order two 
\begin{equation}
\sigma^2(\tilde{O}) \sim O[({\psi_T}-\psi_0)({\psi_T}^\prime-{\psi_0}^\prime)].
\label{eqint4}
\end{equation}

To summarize, using the HF theorem we are able to construct 
an improved observable $\tilde{O}$, Eq.(\ref{tildeobias}), 
having a quadratic zero-variance-zero-bias property, 
Eqs.(\ref{eqint3},\ref{eqint4}), similar to what is known for the energy case, 
Eqs.(\ref{deltaorder},\ref{varorder}). The improved estimator
$\tilde{O}$ depends only on one single quantity, namely ${\psi_T}(\lambda)$. 
Accordingly, to get accurate results we need to choose in the 
neighborhood of $\lambda=0$ a trial function accurate enough to get
not only a small difference in wave functions but also 
in the derivative of the wave functions. In practice, this latter point 
is particularly difficult to fulfill. Indeed, at fixed values of $\lambda$,
it is known that the minimization of the fluctuations of the local energy can allow 
an important reduction of the error in the trial wave function. However, 
there is no reason why it should also lead to a satisfactory representation of
the derivative of the trial wave function.

In order to escape from this difficulty we propose here to 
work directly at $\lambda=0$ and to optimize {\it independently} the 
trial wave function ${\psi_T}$ and its derivative ${\psi_T}^\prime$.
Such procedure is justified since it corresponds to choose as 
$\lambda$-dependent trial wave function the following expression
\begin{equation} 
{\psi_T}(\lambda)= {\psi_T} + \lambda \tilde{\psi},
\end{equation} 
where $\tilde{\psi}$ is some new independent function playing the role 
of a trial function for the derivative of the ground-state at $\lambda=0$.
In this case, the renormalized observable can be rewritten under the
final form
\begin{equation}
\tilde{O} \equiv  O + \frac{(H-E_L)\tilde{\psi}}{\psi_T}
+2 (E_L-E_v) \frac{\tilde{\psi}} {\psi_T}.
\label{tildeobias2}
\end{equation}
where the pair of functions $(\psi_T,\tilde{\psi})$ is the current 
guess for the exact solution $(\psi_0,{\psi_0}^\prime)$.

Let us now turn our attention on the problem of optimizing the two 
trial functions $(\psi_T,\tilde{\psi})$.
Regarding  ${\psi_T}$ we know that the standard procedure consists
in minimizing the variance of the local energy with respect to the 
parameters of the trial function.
Quite remarkably, we have here a similar result for $\tilde{\psi}$:
the best choice is obtained by minimizing the 
variance of the renormalized operator $\tilde{O}$ with respect to 
the parameters of $\tilde{\psi}$.
 
To prove this property it is sufficient to show that 
the zero-variance (or zero-fluctuations) equations for ${E_L}$  
and $\tilde{O}$:
\begin{eqnarray}
{E_L} & = &  \langle {E_L} \rangle_{\psi_T^2} \nonumber \\
 \tilde{O} & = & \langle
\tilde{O} \rangle_{{{\psi_T}}^2}
\label{2var0}
\end{eqnarray}
are equivalent to the equations defining $\psi_0$ and $\psi_0^\prime$,
namely
\begin{eqnarray}
(H-{E_0})\psi_0 & = & 0 \nonumber  \\
(H-{E_0})\psi_0^\prime & + & (O- \langle O \rangle_{\psi_0^2}  )\psi_0
=  0
\label{eqfip}
\end{eqnarray}
In these formulae, the first equation is just the ordinary 
Schroedinger equation. The second one is obtained by deriving 
the Schroedinger equation:
\begin{equation}
H(\lambda)\psi(\lambda)=E_0(\lambda) \psi(\lambda)
\end{equation}
with respect to $\lambda$ at $\lambda=0$.
Note that equations (\ref{eqfip}) determine an unique solution,
($\psi_0,\psi_0^\prime,E_0,\langle O \rangle_{\psi_0^2}$),
as soon as $H$ has a non-degenerate ground-state.
Now, using Eqs.(\ref{tildeobias}) and (\ref{localenergy}) for the 
definition of $\tilde{O}$ and 
$E_L$, respectively, the system of equations (\ref{2var0}) 
can be rewritten under the form
\begin{eqnarray}
(H-{E_v}) {\psi_T} & =  & 0 \\
(H-{E_v}) {\psi_T}^\prime & + & (O-\langle O \rangle_{{{\psi_T}}^2})
 {\psi_T} = 0
\label{eqfippsi}
\end{eqnarray}
which are nothing but Eqs.(\ref{eqfip}) with $({\psi_T},{\psi_T}^\prime)
=(\psi_0,\psi_0^\prime)$. Accordingly, the zero-variance equations (\ref{2var0}) 
admits this latter pair of functions as unique solution.

In practical calculations, different strategies of optimization 
can be employed. A first approach consists in minimizing {\it separately}
the variance of the local energy with respect to the wave function 
${\psi_T}$ and the variance of $\tilde{O}$ with respect
to $\tilde{\psi}$. In this way, we get an optimal trial wave function 
$\psi_T$ for the energy and the best derivative at fixed $\psi_T$. However,
let us emphasize that this approach is not the most general: we can also
minimize both variances simultaneously with respect to the two independent 
functions. Another remark is that the second equation of system (\ref{eqfip}) 
can be viewed as an ordinary first-order perturbation equation.
This is expected since, when $\lambda O$ is considered as a 
perturbation of the Hamiltonian $H$, $\psi_0^\prime$ is nothing but 
the first-order correction to the ground-state 
and $\langle O \rangle_{\psi_0^2}$ the first-order correction 
to the energy.

Finally, let us end this section by commenting in more detail 
the various terms entering expression (\ref{tildeobias2}) of 
the improved operator. Three different contributions can be distinguished:

(i) The ordinary bare estimator $O$ corresponding to $\tilde{\psi}=0$.

(ii) A second contribution given by $(H-{E_L})\tilde{\psi}/{\psi_T}$. 
It is easy to verify that this contribution has a zero average
over the density ${\psi_T}^2$
\begin{equation}
\langle (H-{E_L})\tilde{\psi}/{\psi_T} \rangle_{{{\psi_T}}^2}=0.
\end{equation}
Accordingly, its role is to lower the variance 
of the improved estimator without changing the average of the observable
(no influence on the systematic error). 
Note that for applications where the stationary density is known and can 
be exactly sampled (that is, there is no systematic error in the average) the use of 
contributions (i) and (ii) is sufficient. Important examples 
include all ``classical'' Monte Carlo simulations based on the Metropolis 
algorithm or one of its variants. Such a possibility was the subject of a 
previous work.\cite{zvp1}
(iii) A third term given by $2({E_L}-{E_v})\frac{\tilde{\psi_T}}{{\psi_T}}$.
This contribution has a very small impact on the statistical
fluctuations since the variance of $({E_L}-{E_v})$ is of order 
two in the trial wave function error for any choice of $\tilde{\psi}$. 
Its main effect 
is to take into account the change of stationary density under the external 
perturbation defined by the observable 
and, therefore, to lower the systematic error in the expectation value of the observable.
Note that in the limit ${\psi_T}=\psi_0$, 
this contribution reduces to zero and, therefore, the average of this term can be 
understood as a correction to the Hellmann-Feynman formula
when ${\psi_T}$ is not the exact ground-state (note that similar corrections to the HF formula 
exist also in more traditional {\it ab initio} calculations, e.g. the
``Pulay force''\cite{pulay} resulting from approximate Hartree-Fock (or LDA) orbitals 
in self-consistent schemes).

\subsection{More improved estimators: use of coordinate transformations}

In this section it is shown how to generalize further our 
renormalized operators.
The basic idea of the generalization is based on an original idea 
recently proposed by Filippi and Umrigar in their work 
on the computation of forces \cite{fil3}. Working in a finite difference formalism 
they have proposed to compute the forces as a small but finite difference
of energies for two close enough geometries. In order to minimize the 
fluctuations they have proposed to use a correlated sampling method in 
which a common Monte Carlo density (the so-called primary one) 
is used for the two close geometries. 
Written within our notations and taking the limit of the two geometries 
infinitely close ($\delta R \rightarrow$ 0 is equivalent to $\lambda \rightarrow$ 0) it means that 
the variational energy is written under the form
\begin{equation}
E_v(\lambda)=
\frac{\langle {E_L}(\lambda) \frac{{{\psi_T}^2(\lambda)}}{\psi_T^2} 
\rangle_{\psi_T^2}}
{\langle \frac{{{\psi_T}^2(\lambda)}}{\psi_T^2} \rangle_{\psi_T^2}}
\end{equation}
where ${\psi_T}(\lambda)$ is the trial wave function chosen for a parameter
$\lambda$ and $\psi_T$ is the reference (primary) trial wave function. 

The price to pay when doing that is the introduction of 
some additional fluctuations associated with the weight
$\frac{{{\psi_T}^2(\lambda})}{\psi_T^2}$.
The remedy they propose to deal with this problem is to use  
a specific coordinate transformation (space-warp transformation)
based on physical motivations: The transformation is built 
so that the electrons close to a given nucleus move almost rigidly with 
that nucleus when the geometry is changed. 
Here, we generalize this idea: coordinate 
transformation can help to minimize the relative fluctuations when 
varying the external parameter $\lambda$. As a physical consequence, estimators 
built from the derivative are expected to have smaller fluctuations 
and smaller systematic errors. 

Let us write a general coordinate transformation as follows
\begin{equation}    
\vec{y}= \vec{T}(\lambda,\vec{x})
\end{equation}  
where the vector $\vec{x}$ (or $\vec{y}$) denotes the set of 
the 3$n_{elec}$ electronic coordinates. Using this transformation the variational energy 
at a given $\lambda$ can be written as
\begin{equation}
E_v(\lambda)=
\frac{\langle 
E_L[\lambda,\vec{T}(\lambda,\vec{x})] 
J(\lambda,\vec{x})
\frac{ {\psi_T}^2[\lambda,\vec{T}(\lambda,\vec{x})] }
     { {\psi_T}^2(\vec{x}) }
\rangle_{{\psi_T}^2} }
{
\langle 
J(\lambda,\vec{x}) 
\frac{{\psi_T}^2[\lambda,\vec{T}(\lambda,\vec{x})]}
     {{\psi_T}^2(\vec{x})} \rangle_{{\psi_T}^2}
}
\label{eJacprime}
\end{equation}
where $J(\lambda,\vec{x})$ is the Jacobian of the transformation. 
Introducing the vector field ${\vec{v}}$ such that at first order
in $\lambda$ we have:
\begin{equation}
\vec{T}(\lambda,\vec{x})=\vec{x}+\lambda \vec{v}(\vec{x})+O(\lambda^2)
\end{equation}
we can compute the derivative of the variational energy with respect to 
$\lambda$ at $\lambda=0$. After some simple but tedious algebra we get 
the following equality
\begin{equation}
\frac{d{E_v}(\lambda)}{d\lambda} \big|_{\lambda=0}
= \langle \tilde{O} \rangle_{{\psi_T}^2}
\label{evprime2}
\end{equation}
where $\tilde{O}$ is a new renormalized operator given by
\begin{equation}
\tilde{O} \equiv  O + \frac{(H-{E_L}){\psi_T}^\prime}{{\psi_T}}
+2 ({E_L}-{E_v}) \frac{{{\psi_T}}^\prime}
{{\psi_T}}
+ 
\frac{\vec{\nabla} [(E_L -E_v) \psi_T^2 \vec{v}]}
{\psi_T^2}.
\label{tildeobias3}
\end{equation}
To derive this expression we have used the fact that the Jacobian
defined as
\begin{equation}
J(\lambda,\vec{x})=det[ \frac{\partial T_i(\lambda,\vec{x})  }{\partial x_j}]
\label{jacobian}
\end{equation}
has the following small-$\lambda$ expression
\begin{equation}
J(\lambda,\vec{x})=det[ \delta_{ij} + \lambda \frac{\partial v_i}{\partial x_j}]
+ O(\lambda^2)
\label{jacobian2}
\end{equation}
and, therefore,
\begin{equation}
J(0,\vec{x})=1
\end{equation}
\begin{equation}
\frac{\partial J}{\partial \lambda}(0,\vec{x})= \vec{\nabla}.\vec{v}
\end{equation}
This more general operator is identical to the operator derived in the previous 
section plus a new contribution resulting from the derivative of the 
coordinate transformation. This new term has a zero average over
the VMC distribution $\psi_T^2$. Accordingly, its main role is 
to reduce further the statistical error. However, it is important to emphasize that, 
when the trial function 
$\tilde{\psi}$ and the vector field $\vec{v}$ are optimized simultaneously 
it has also an influence on the magnitude of the systematic error.

\section{Beyond Variational Monte Carlo}

In the preceding section we have shown how to construct improved observables,
$\tilde{O}$, associated with accurate expectation values:
\begin{equation}
\frac {\langle \psi_T |\tilde{O}| \psi_T \rangle}
                 { \langle \psi_T | \psi_T \rangle}
=\frac {\langle \psi_0 |O| \psi_0 \rangle}
                 { \langle \psi_0 | \psi_0 \rangle} 
 + O[({\psi_T}-\psi_0)
(\tilde{\psi}-{\psi_0}^\prime)].
\label{bvmc1}
\end{equation}
When the error $\delta \psi^\prime=\tilde{\psi}-{\psi_0}^\prime$ in the trial function for the derivative
is comparable to the error in the trial function for the ground-state, 
$\delta \psi={\psi}_T -{\psi_0}$, the accuracy reached with the preceding variational 
estimate, Eq.(\ref{bvmc1}), can be comparable to the very good accuracy usually obtained for 
total energies. However, despite this remarkable improvment,
we are still left with some small residual systematic error 
associated with approximate $\psi_T$ and $\tilde{\psi}$. In the energy case
it is known that this error can be entirely suppressed (at least for systems with no nodes or 
known nodes) by averaging the local 
energy over the mixed Diffusion Monte Carlo (DMC) probability distribution, 
$\pi_{DMC} \sim \psi_T \psi_0$ instead of the VMC distribution, $\pi_{VMC} \sim {\psi_T}^2$. 
Unfortunately, we have no such result for the improved 
observables defined here. However, as we shall see now, we can still define some 
approximate way for recovering most of the error.

A natural way of defining an exact extimator for the observable is to consider 
the derivative of the exact DMC energy estimator instead of the VMC one
\begin{equation}
{E_0}(\lambda) \equiv \langle {E_L}(\lambda) \rangle_{{\psi_T}(\lambda){\psi_0}
(\lambda) } =
\langle \frac{H(\lambda) {\psi_T}(\lambda)}{{\psi_T}(\lambda)} 
\rangle_{{\psi_T}(\lambda) {\psi_0}(\lambda)}
\end{equation}
Making the derivative and rewritting the result as an ordinary average we get:
\begin{equation}
\frac{d{E_0}(\lambda)}{d\lambda} \big|_{\lambda=0}
= \langle \tilde{O}\rangle_{{\psi_T}{\psi_0}}
\label{e0vprime}
\end{equation}
where $\tilde{O}$ is written as
\begin{equation}
\tilde{O} \equiv  O + \frac{(H-{E_L}){\psi_T}^\prime}{{\psi_T}}
+({E_L}-{E_0}) (\frac{{{\psi_T}}^\prime} {{\psi_T}}
+ \frac{{{\psi_0}}^\prime} {{\psi_0}}).
\label{tildeobias4}
\end{equation}
Of course, written under the above form, this exact estimator is useless since the exact 
wave function is not known. Here, we propose to make the following natural approximation
\begin{equation}
\frac{{{\psi_0}}^\prime} {{\psi_0}} = \frac{{{\psi_T}}^\prime} {{\psi_T}}.
\label{approxdmc}
\end{equation}
Therefore, our final approximate DMC estimator is written as
\begin{equation}
\tilde{O}_{DMC} \equiv  O + \frac{(H-{E_L})\tilde{\psi}}{{\psi_T}}
+2 ({E_L}-{E_0}) \frac{{\tilde{\psi}}} {{\psi_T}}
\label{tildeobias5}
\end{equation}
where $\tilde{\psi}$ is as usual our trial function for the derivative of 
the exact wave function. Note that this estimator is very similar to the VMC one, Eq.(\ref{tildeobias2}).
The only difference lies in the value of the average energy, $E_0 =\langle E_L \rangle$, 
entering the definition of $\tilde{O}_{DMC}$. More precisely, we have
\begin{equation}
\tilde{O}_{DMC}- \tilde{O}_{VMC} = 2 (E_v -E_0) \frac{{\tilde{\psi}}} {{\psi_T}}.
\label{f10}
\end{equation}

Now, in order to reduce further the error let us show that we can generalize the 
usual ``hybrid formula'' ${\langle O \rangle}_{hybrid} \equiv 2{\langle O \rangle }_{DMC}
- {\langle O \rangle }_{VMC}$ defined for bare observables to the case of improved observables. 
To do that, let us develop the quantity $\langle \delta \psi | \tilde{O}_{DMC}
 |\delta \psi \rangle$
where $\delta \psi= \psi_T - \psi_0$
\begin{equation}
\langle \delta \psi | \tilde{O}_{DMC}|\delta \psi \rangle = 
\langle \psi_T | \tilde{O}_{DMC}|\psi_T \rangle 
-2 \langle \psi_T | \tilde{O}_{DMC}|\psi_0 \rangle
+ \langle \psi_0 | \tilde{O}_{DMC}|\psi_0 \rangle
\end{equation}
which leads to
$$
2 \langle \psi_T | \tilde{O}_{DMC}|\psi_0 \rangle
-\langle \psi_T | \tilde{O}_{DMC}|\psi_T \rangle 
= \langle \psi_0 | O |\psi_0 \rangle
 + A + O( {\delta \psi}^2)
$$
where the intermediate quantity $A$ is defined as
$$
A \equiv 
\langle \psi_0 | \frac{(H-{E_L})\tilde{\psi}}{{\psi_T}}
+2 ({E_L}-{E_0}) \frac{{\tilde{\psi}}} {{\psi_T}} | \psi_0 \rangle
$$
Expanding $A$ in terms of $\delta \psi= \psi_T-\psi_0$ we get 
$$
A = 2 \langle \psi_T |E_L - E_0|\tilde{\psi} \rangle 
   -2 \langle  \delta \psi| (H-E_L) |\tilde{\psi}\rangle 
-4 \langle \delta \psi |E_L - E_0|\tilde{\psi} \rangle 
 + O( {\delta \psi}^2)
$$
Using now the equality
$$
E_L-E_0= \frac{ (H -E_0)\delta \psi} {\psi_T}
$$
we obtain
$$
A = O( {\delta \psi}^2)
$$
This latter result shows that the error in the hybrid estimator is 
of order two in $\delta \psi$
\begin{equation}
2 \langle \tilde{O}_{DMC} \rangle_{\psi_T  \psi_0} - 
  \langle \tilde{O}_{DMC} \rangle_{\psi_T^2} = \langle \psi_0 | O |\psi_0 \rangle + O( {\delta \psi}^2),
\label{hybridfinal}
\end{equation}
thus generalizing the standard result for the bare observable.
Note that we can use either $\tilde{O}_{DMC}$ or $\tilde{O}_{VMC}$ 
[Eq.(\ref{tildeobias2})] in this latter 
formula since the difference between the two renormalized operators is proportional to $E_v-E_0$, Eq.(\ref{f10}), 
and, therefore, is also of order two in $\delta \psi$ [Eq.(\ref{deltaorder})].

When using coordinate transformation we have similar results. The 
exact DMC estimator is found to be
\begin{equation}
\tilde{O} \equiv  O + \frac{(H-{E_L}){\psi_T}^\prime}{{\psi_T}}
+({E_L}-{E_0}) (\frac{{{\psi_T}}^\prime} {{\psi_T}}
+ \frac{{{\psi_0}}^\prime} {{\psi_0}})
+
\frac{\vec{\nabla} [(E_L -E_0) \psi_T \psi_0 \vec{v}]}
{\psi_T \psi_0}
\label{tildeobias6}
\end{equation}
and we propose to use the following approximate form
\begin{equation}
\tilde{O}_{DMC} \equiv  O + \frac{(H-E_L)\tilde{\psi}}{{\psi_T}}
+2 ({E_L}-\langle E_L \rangle ) \frac{{\tilde{\psi}}} {{\psi_T}}
+
\frac{\vec{\nabla} [(E_L - \langle E_L \rangle ) {\psi_T}^2 \vec{v}]}
{{\psi_T}^2}
\label{tildeobias7}
\end{equation}
Because the difference $(E_L -\langle E_L \rangle )$ is of order $\delta \psi$ it is easy to 
verify that the error in the hydrid estimator given by Eq.(\ref{hybridfinal}) 
remains here also of order two.

Before ending this section let us emphasize that it is possible to write a closed 
computable expression for the exact estimator of the observable,  
Eq.(\ref{tildeobias4}), by expressing the unknown quantity 
$ \frac{{{\psi_0}}^\prime} {{\psi_0}}$ as a computable 
stochastic average. Choosing a $\lambda$-independent trial 
wave function $\psi_T$ we can write\cite{ceperley2,cafstat,caffaclav}
\begin{equation}
\psi_0(\lambda,x) = \psi_T(x)  \lim_{T \rightarrow +\infty}
<< e^{-\int_0^{T} ds E_L[\lambda,x(s)]} >>_{x(0)=x}
\label{psiPDMC}
\end{equation}
where $x$ denotes an arbitrary point in configuration space and 
$<<...>>_{x(0)=x}$ denotes the sum over all drifted random walks of length $T$ starting 
at $x$ as obtained in a Pure Diffusion Monte Carlo (PDMC) scheme 
(DMC without branching).\cite{caffaclav} 
Of course, a similar formula can also be obtained in a DMC scheme.\cite{liu1} 
Now, using formula (\ref{psiPDMC}) we get
\begin{equation}
\frac{{{\psi_0}}^\prime} {{\psi_0}}= \lim_{T \rightarrow +\infty}
\int_0^{T} dt \frac{
<< O[x(t)] e^{-\int_0^{T} ds E_L[\lambda,x(s)]}>>_{x(0)=x} }
{<< e^{-\int_0^{T} ds E_L[\lambda,x(s)]}>>_{x(0)=x} }
\end{equation}
and, therefore, the exact estimator can be written in terms of a standard part plus 
a time-integral of the two-point correlation function between the local energy 
and the observable
$$
\tilde{O} = O + \frac{(H-{E_L})\tilde{\psi}}{\psi_T}
+(E_L-\langle E_L \rangle) \frac{{\tilde{\psi}}} {\psi_T}
$$
\begin{equation}
+ \lim_{T \rightarrow +\infty}
\int_0^{T} dt  \frac{
<< (E_L- \langle E_L \rangle )[x(0)] O[x(t)] e^{-\int_0^{T} ds E_L}>>_{x(0)=x} }
{<< e^{-\int_0^{T} ds E_L}>>_{x(0)=x} }
\end{equation}
It is important to emphasize that this latter estimator is exact: averaged 
over the mixed DMC distribution it leads to an unbiased estimate of the exact 
average. However, the correlator can only obtained within a Forward Walking 
scheme and, therefore, the stability in time is not guaranteed. In this work, we shall 
not use this expression, its implementation will be presented in a forthcoming work.

\section{Application to forces}

The average force between atoms in a molecular system is defined as
\begin{equation}
\bar{F}_{q_i} \equiv -\frac{\partial E_0({\bf q})}{ \partial q_i},
\label{force}
\end{equation}
where $E_0({\bf q})$ is the total electronic ground-state energy for a given nuclear configuration;
${\bf q}$ represents the $3N_{nucl}$ nuclear coordinates ($N_{nucl}$, number of nuclei) and
$q_i$ the particular force component in which we are interested.

Defining the local force as follows
\begin{equation}
F_{q_i}({\bf x}, {\bf q}) \equiv - \frac{\partial V({\bf x},{\bf q})}{ \partial q_i}.
\label{localforces}
\end{equation}
where ${\bf x}$ represents the $3n_{elec}$ electronic coordinates ($n_{elec}$, number of electrons)
and $V$ the total potential energy operator, and making use of the
Hellmann-Feynman (HF) theorem the average force can be rewritten as
the statistical average of the local force over the exact distribution $\psi_0^2({\bf x})$:
\begin{equation}
\bar{F}_{q_i} = \langle F_{q_i}(\bf x, q) \rangle_{\psi_0^2({\bf x})}.
\end{equation}
Written under this form the various proposals presented in the preceding sections can be applied 
to the calculation of the average force.
It is important to emphasize that for approximate probability densities (VMC or DMC) the HF theorem is no 
longer valid and a systematic error in the statistical average $\langle F_{q_i}(\bf x, q)\rangle$ is introduced.
However, it is not a problem here since it is the 
purpose of this work to show that, by using suitable improved estimators, this error 
can be reduced and even suppressed in the zero-bias limit. 

In order to discuss the various aspects of the method we shall restrict ourselves to the 
case of diatomic molecules. Let us consider a diatomic molecule $AB$ with atom $A$ located 
at $(R,0,0)$ and atom $B$ located at the origin. The only non-zero component of the local force 
acting on the nucleus $A$ is the $x$-component given by
\begin{equation}
F = - \frac{\partial V}{\partial R}
= \frac{Z_A Z_B}{R^2} - Z_A \sum_{i=1}^{n_{elec}} \frac{(x_i-R)}{|{\bf r_i -R}|^3}.
\label{localforcediatomics}
\end{equation}

In this work we present a number of VMC and fixed-node DMC calculations for the diatomic molecules 
H$_2$,LiH, and Li$_2$. Implementation of the quantum Monte Carlo methods is well-known and will not be
discussed here. For the H$_2$ molecule the trial wave function used has the following simple form
\begin{equation}
\psi_T= (1s_A 1s_B + 1s_B 1s_A) + c (1s_A 1s_A + 1s_B 1s_B)
\end{equation}
where $1s_M$ is a 1s-Slater function centered at nucleus $M=A,B$ with 
exponent $\mu$ and $c$ a parameter describing the amount of ionic contribution 
into the wave function. Of course, much more accurate trial wave functions can be constructed for H$_2$. However, 
our purpose here is to show that such a simple form for $\psi_T$ is already sufficient to 
get accurate values of the force.

For LiH and Li$_2$ we have employed two types of trial wave function. Our main choice
is standard in QMC calculations for molecules. The trial 
wave function is made of a determinant of single-particle orbitals multiplied by a Jastrow 
factor. The determinantal part is obtained from a RHF calculation 
and only the Jastrow factor is optimized.
As we shall see below, we have also used Valence-Bond (VB)-type wave functions
consisting of a number of determinants multiplied by a Jastrow factor. We have used such 
a multideterminantal description to reproduce correctly 
the large interatomic distance regime (dissociation limit). In the case of LiH 
the determinantal part consists of three determinants corresponding to 
the covalent VB resonating structure: $(1s_{Li})^2[ 2s_{Li} 1s_H + 1s_H 2s_{Li}]$ ($\{1s_{Li},2s_{Li},1s_H\}$ 
optimized {\it atomic} orbitals for the Li and H atoms, orbitals occupied 
by electrons $\alpha$ and $\beta$ antisymmetrized separately) 
and one ionic VB structure: $(1\tilde{s}_{Li})^2 1\tilde{s}_H 1\tilde{s}_H$ 
($1\tilde{s}_{Li},1\tilde{s}_H$ optimized atomic orbitals for the Li$^+$ and H$^-$ ions). 
In the case of Li$_2$ we have considered a six-determinant 
representation consisting of 
the three covalent VB structures describing the resonance between atomic orbitals $(2s_A,2s_B)$, 
$(2p_{y_A},2p_{y_B})$, and $(2p_{z_A},2p_{z_B})$. 
This latter trial wave function reproduces not only the dissociation limit 
but also a major part of the 2s-2p near-degeneracy.

In Figures \ref{figureEH2},\ref{figureELiH}, and \ref{figureELi2}
the energy curves obtained for H$_2$,LiH, and Li$_2$ are presented.
Upper curves are the VMC curves (open squares joined by a dotted line). 
For H$_2$ the two parameters $c$ and $\mu$ have been optimized for each 
interatomic distance. For LiH and Li$_2$ the Jastrow-RHF wave function (one determinant) has been used.
All the parameters entering the Jastrow factor have been optimized for all distances.
Optimizations have been performed by minimizing the variance of the local energy 
using the correlated sampling method of Umrigar {\it et al.}\cite{umrigar}. 
The first important observation is that, except for H$_2$, VMC curves are not 
smooth as a function of $R$. Such a result is not surprising: 
It is typical of a situation where an approximate trial wave function 
is optimized {\it independently} for different 
values of an external parameter (here, $R$) with respect 
to a large number of variables (for LiH and Li$_2$ we have used about 30 independent 
variational parameters). Depending on the initial conditions (which are themselves very dependent on $R$)
the algorithm used for minimizing the variance can be trapped within one of the various {\it local}
minima. As a consequence, the actual value obtained for the variance
(and the corresponding energy) can vary abruptly even when the 
external parameter is changed smoothly. 
Of course, this problem can be solved in principle by making very careful optimizations on very large samples.
Indeed, the functional form of the trial wave function being identical 
at all distances a smooth curve must be got when the correct lowest minimum 
of the variance is obtained at each distance. Here, this is the case for
H$_2$ whose trial wave function contains only two variational parameters.
However, for large systems including a much larger number 
of variational parameters and nuclear degrees of freedom, the possibility of fully optimizing 
the trial wave function is just irrealistic.
As an important consequence, let us emphasize
that, in practice, there is no hope of 
obtaining meaningful forces by making straight finite differences of optimized variational energies
without using some sort of correlated sampling scheme. This is a good illustration of how difficult the 
calculation of forces is within a QMC framework. 

Intermediate points (filled squares) are the DMC results obtained from
fixed-node calculations using the optimized VMC trial wave functions. 
In sharp contrast with VMC, the DMC curves are now regular. 
This is so because, unlike VMC, fixed-node DMC averages 
do not depend on the particular form of the trial function used, except for 
the nodal structure. Here, the nodal hypersurfaces vary smoothly as a function 
of the distance and, therefore, the corresponding DMC energy curves are smooth 
within error bars.
The second important observation is related to the global shape of the curves.
Ideally, we are interested in having energy curves which differ from the exact curve 
only by a constant independent on the distance. This is indeed the condition 
to obtain accurate derivatives. Here, it is not the case for the 
VMC curves of LiH and Li$_2$: the difference between the VMC and  
exact energy curves is an increasing function of the distance. 
It is not a surprise since in both cases the trial wavefunction is built 
from a single RHF determinant based on delocalized molecular orbitals which leads to a 
wrong description of the dissociation 
limit. However and, very interestingly, fixed-node DMC results have a much better 
behavior at large distances. As a consequence, one may expect at this stage to obtain 
accurate forces from the derivative of the fixed-node energy curve even when 
relatively crude wave functions are used.
Finally, let us note that the quality of the fixed-node calculations for the 
molecules considered here is quite good. To give an example, at the equilibrium 
distance of the Li$_2$ molecule, the total energy obtained is -14.9901(6) 
to be compared with the exact non-relativistic value of $E_0$=-14.9954. 
The amount of correlation energy recovered within the fixed-node approximation 
is about $95.7\%$. A similar quality is obtained for other 
distances and also for the LiH molecule. 
In the case of the nodeless H$_2$ molecule (no fixed-node approximation), the DMC 
energies agree perfectly well with the exact ones.
For the H$_2$ molecule the variance of the local energy varies between 0.3 at $R=0.8$ and 
0.02 at $R=3.5$; for LiH the variance is about 0.07, and for Li$_2$ it varies between 0.09 and 0.2.

The crucial point when implementing the various formulae presented in the preceding 
section is the choice of the trial function $\tilde{\psi}$ for the derivative.
In our previous study on forces\cite{zvp2} 
where we have focused our attention on
the reduction of statistical fluctuations only, we have proposed to employ
the minimal form leading to a finite variance of the renormalized local force.
As can be viewed from Eq.(\ref{localforcediatomics}), at short electron-nucleus 
distance $r$ the local force behaves as $F\sim 1/r^2$ and, therefore, the 
variance $ \langle F^2 \rangle -{\langle F \rangle }^2$ is infinite. This  
well-known problem has been discussed in different places 
(See, {\it e.g.} \cite{lester2} Chap. 8.2 or \cite{vrbik2}). Here, 
the ``minimal'' form removing the singular part responsible for the infinite 
variance is written as
\begin{equation}
\tilde{\psi}_{min}({\bf x})= Q {\psi_T}
\label{psi}
\end{equation}
where $Q$ is given by
\be
Q=  Z_A \sum_{i=1}^{n_{elec}} \frac{(x_i -R)} { {|{\bf r}_i
 - {\bf R}|}  }.
\label{q}
\ee
To see this, we just need to compute the following quantity
\be
\frac{ (H-E_L) \tilde{\psi}_{min} } {\psi_T}
= Z_A \sum_{i=1}^{n_{elec}} \frac{(x_i-R)}{|{\bf r_i -R}|^3}
 - {\bf \nabla} Q \cdot {\bf \nabla}\psi_T /\psi_T.
\ee
By adding this latter quantity to the bare local force, Eq.(\ref{localforcediatomics}),
the singular part is exactly removed, the remaining contribution 
having a finite variance. In what follows, $\tilde{\psi}_{min}$ will be referred to as
the minimal form for $\tilde{\psi}$.

In Figure \ref{forcevar} we present various VMC calculations of the average 
force for the Li$_2$ molecule as a function of the interatomic distance.
A first set of points (filled squares points with very large error bars at $R$=5.,$R$=6.5, and $R$=7.5) are results 
obtained from the ordinary bare estimator, Eq.(\ref{localforcediatomics}).
Open squares (with small error bars) joined by the dashed curve  correspond to results obtained by using 
${\tilde{\psi}_{min}}$ as trial function for the derivative
\begin{equation}
\tilde{F}_{VMC-ZV}[\psi_T,\tilde{\psi}_{min}] 
= F + \frac{(H-{E_L}) {\tilde{\psi}_{min}}}{{\psi_T}}.
\label{f1}
\end{equation}
This estimator can be viewed as the simplest improved estimator we can think of 
having a finite variance; it corresponds to the form employed in our previous study.\cite{zvp2}
The subscript $ZV$ (Zero-Variance) is used here to 
emphasize that the improved estimator is built to decrease the statistical 
error only. Circles joined by a dotted line are results obtained from
the ZVZB improved estimator derived in the preceding section, Eq.(\ref{tildeobias2}):
\begin{equation}
\tilde{F}_{VMC-ZVZB}[\psi_T,\tilde{\psi}_{min}]
= F + \frac{(H-{E_L}) {\tilde{\psi}_{min}}}{{\psi_T}} + 2(E_L- \langle E_L \rangle )
\frac{\tilde{\psi}_{min}}
{\psi_T}.
\label{f2}
\end{equation}
Note the use of the subscript $ZVZB$ to emphasize on the two aspects: reduction of statistical and 
systematic errors. Finally, the solid line represents the ``exact''
non-relativistic force curve for Li$_2$.

A first important observation is that using improved estimators 
is extremely efficient in reducing the statistical error. This can be seen by comparing 
the magnitude of the error bars on data obtained from the ordinary bare estimator 
(filled squares at $R=5.,6.5$, and 7.5) with those corresponding to other calculations based 
on improved estimators. A reduction of at least two orders of magnitude is observed.
As already discussed this remarkable result is a direct consequence of the fact that 
the infinite variance of the bare estimator has been reduced to a finite value.
At the scale of the figure error bars associated with improved estimators are almost not visible. 
A more quantitative analysis will be given later (see, Table \ref{tableLi2}).

Now, regarding systematic errors, results are much more disappointing. Using 
the pure Zero-Variance (ZV) renormalized estimator, Eq.(\ref{f1}), the behavior of the average 
force (open squares joined by the dashed line in Fig.\ref{forcevar}) as a function of $R$ appears erratic. 
This can be easily understood since the term added 
to the bare force in Eq.(\ref{f1}) has a zero-average and, therefore, the erratic behavior is a direct 
consequence of the irregular VMC energy curve presented in Figure \ref{figureELi2}.
When adding the term correcting the average, the 
results are improved. As seen on the figure the behavior of 
$\langle \tilde{F}_{VMC-ZVZB}[\psi_T,\tilde{\psi}_{min}]\rangle$, Eq.(\ref{f2}), as a function of $R$, 
is much less irregular, thus illustrating the important role played by the zero-bias
additional contribution [third term of the R.H.S. of Eq.(\ref{f2})] to correct the error due to the approximate trial wave function.
Despite of that, the resulting curve is far from being satisfactory.
To weaken the role played by $\psi_T$ we can think of going 
beyond VMC calculations. In Figure \ref{forcedmc}
we present such calculations for Li$_2$ using the DMC-ZVZB improved estimator 
$\tilde{F}_{DMC-ZVZB}[\psi_T,\tilde{\psi}_{min}]$ written as
\begin{equation}
\tilde{F}_{DMC-ZVZB}[\psi_T,\tilde{\psi}_{min}]
= F + \frac{(H-E_L) {\tilde{\psi}_{min}}}{{\psi_T}} + 2(E_L- \langle E_L \rangle)
\frac{\tilde{\psi}_{min}}
{\psi_T}
\label{f3}
\end{equation}
where the energy average is a fixed-node DMC average,
and, also, results obtained by using the generalized hybrid formula, 
Eq.(\ref{hybridfinal})
\begin{equation}
\bar{F} \simeq 2 \langle \tilde{F}_{DMC-ZVZB} \rangle_{\psi_T  \psi_0} -
\langle \tilde{F}_{VMC-ZVZB} \rangle_{\psi_T^2} 
\label{hybridfinal2}
\end{equation}
A clear improvment is observed when going from VMC (open circles) to DMC (filled squares) and, then, 
to hybrid calculations (open squares): The systematic error present in VMC calculations is reduced. 
However, the resulting curves are still not satisfactory. Extracting from them a meaningful equilibrium 
distance or first derivative of the force curve (calculation of $\omega_e$) is impossible.
Very similar behaviors have been obtained for H$_2$ and LiH. They do not need to be reproduced here.

The main reason for the poor results just presented is the 
low quality of the trial function $\tilde{\psi}_{min}$ used. According to our 
general presentation of Sec. II we know that a good trial function $\tilde{\psi}$ must 
be close to the derivative of the exact ground-state wave function with respect to $R$.
Here, this is only true when an electron approaches the nucleus A (note that nucleus B has been fixed  
at the origin and, thus, has no pathological contribution). In that case the non-vanishing part of the 
exact wave function is expected to behave as
\begin{equation}
\psi_0 \sim_{ {\bf r}_i \rightarrow {\bf R}} \exp{(-Z_A | {\bf r}_i- {\bf R}|)}
\end{equation}
which leads to 
\begin{equation} 
\frac{\partial \psi_0}{\partial R} \sim_{ {\bf r}_i \rightarrow {\bf R} }
-Z_A \frac{ (x_i -R)} { {|{\bf r}_i - {\bf R}|} } \psi_0
\end{equation}
which is nothing but (up to a minus sign) the minimal form for $\tilde{\psi}$ given above, Eqs.(\ref{psi},\ref{q}).

In order to improve our trial wave function $\tilde{\psi}$ we propose to use 
the following finite-difference form 

\begin{equation}   
\tilde{\psi}_{Deriv}= \frac{\psi_T[R+\Delta R, p(R+\Delta R)]- 
\psi_T[R, p(R)] } { \Delta R }.
\label{psideriv}
\end{equation}
In this expression $p(R)$ denotes the complete set of variational parameters entering 
the trial wave function (coefficients of molecular orbitals, basis set exponents, 
Jastrow parameters, etc...). The main advantage of using a finite-difference form 
instead of the exact derivative is practical:  To estimate 
the derivative we only need to compute two additional local energies and, thus, we 
avoid deriving and programming the lengthy expressions resulting 
from the explicit derivative. Note also that  using an approximate
finite-difference representation is not a problem here:
In any case $\tilde{\psi}_{Deriv}$ must be considered 
as an approximate trial function for the exact derivative and $\Delta R$ 
can always be interpreted as a new additional variational parameter for $\tilde{\psi}$.
In practical calculations, the complete set of parameters we use for 
minimizing the fluctuations of the various improved estimators
consists of $ \{ p(R)$, $p(R+\Delta R)$, and $\Delta R \}$.

At the VMC level we consider the following form for the improved estimator
$$
\tilde{F}_{VMC-ZVZB}[\psi_T,\tilde{\psi}_{Deriv},\vec{v}]= 
$$
\begin{equation}
F+ \frac{(H-E_L) \tilde{\psi}_{Deriv}}{\psi_T} 
+ 2(E_L - \langle E_L \rangle ) \frac{\tilde{\psi}_{Deriv}}{\psi_T}
+
\frac{\vec{\nabla} [(E_L - \langle E_L \rangle) \psi_T^2 \vec{v}]}
{\psi_T^2}.
\label{fpsir}
\end{equation}
At the DMC level the expression used is very similar, see Eq.(\ref{tildeobias7}).
In this expression the vector field $\vec{v}$ associated with the coordinate
transformation is chosen as follows
\begin{equation}
\vec{v}=\sum_{i=1}^{n_{elec}} e^{-\alpha r_{iA}-\beta r_{iA}^2} \vec{u_x}
\end{equation}
where $\vec{u_x}$ is the unit vector along the x-axis.
The vector field depends on two parameters $\alpha$ and $\beta$, which are
optimized to lower the variance of $\tilde{F}_{VMC-ZVZB}$.
The vector field is built so that electrons close to the nucleus A translate with the 
nucleus, while electrons far away do not move.
In practice, we compute the additional term associated with the coordinate 
transformation using a finite-difference scheme along the direction defined by the 
vector $\vec{v}$
$$
\frac{\vec{\nabla} [(E_L - \langle E_L \rangle ) \psi_T^2 \vec{v}]}{\psi_T^2} = 
$$
\begin{equation}
(E_L - \langle E_L \rangle) \nabla . {\vec{v}} + 
[(E_L- \langle E_L \rangle)(\vec{x}+\epsilon \vec{v}) \frac{\psi_T^2( \vec{x}+\epsilon \vec{v})}
{\psi_T^2 (\vec{x})} - (E_L- \langle E_L \rangle)(\vec{x})]/ \epsilon
\label{bestestimator}
\end{equation}
where $\vec{x}$ represents the electronic coordinates and  $\epsilon$ a small 
positive quantity whose magnitude can also be optimized.

In Figure \ref{fLi2psiR} we present VMC calculations 
for Li$_2$ using 
$\tilde{F}_{VMC-ZVZB}[\psi_T,\tilde{\psi}_{min}]$ and 
$\tilde{F}_{VMC-ZVZB}[\psi_T,\tilde{\psi}_{Deriv},\vec{v}]$ as 
improved estimators.  The two estimators have been evaluated on the same Monte Carlo samples.
There are two striking differences when using the second estimator 
$\tilde{F}_{VMC-ZVZB}[\psi_T,\tilde{\psi}_{Deriv},\vec{v}]$.
First, the gain in statistical error is spectacular (about one order of magnitude, for a quantitative analysis see 
discussion below, Table \ref{tableLi2}). Second, the curve is much more regular and closer to the exact 
result (solid line). The VMC results are very satisfactory at small distances 
(between R=3. and R=4.). However, at larger interatomic distances, the VMC curve 
begins to separate from the exact one. This is due to the fact that the wave function 
is built from a RHF calculation and, therefore, the dissociation limit 
is not correctly described. To address this problem we have considered a more 
sophisticated trial wave function consisting of a product of a Jastrow factor 
and a six-determinant one-particle part (for a more detailed description, see above).
This VB-type trial wave function has been 
used only for the largest distances: R=7.,R=7.5,R=8, and R=8.5
In figures \ref{fLi2finalvmc} and \ref{fLi2finaldmc} the comparison between results 
obtained with the Jastrow-RHF (one determinant) and the Jastrow-VB (six determinants) wave functions 
is presented. At the VMC level (Fig.\ref{fLi2finalvmc}), the improvment resulting from the multideterminant 
wave function is clearly seen, the forces computed are much closer to the exact curve than in the one-determinant 
case. At the DMC level (Fig.\ref{fLi2finaldmc}) we could expect that this error disappears even with 
the Jastrow-RHF (one determinant) wave function since the DMC results depend only on the nodal structure of 
the wave function. 
However, it is not true. The difference between the 
DMC curve and the exact one is still important at large distances like in the VMC case. 
This result takes its origin in the approximation made for the exact derivative of 
the wave function in the DMC estimator, Eq.(\ref{approxdmc})
($\frac{{{\psi_0}}^\prime} {{\psi_0}}$ is replaced by $\frac{{{\psi_T}}^\prime} {{\psi_T}}$).
When using the Jastrow-VB wave function the DMC results obtained are much better.

We are now in a position to present our final curves for the three molecules obtained 
with our best fully-optimized estimator $\tilde{F}_{ZVZB}[\psi_T,\tilde{\psi}_{Deriv},
\vec{v}]$ and the hybrid formula. Results for the molecules H$_2$,LiH, and Li$_2$
are presented in Figures \ref{fH2final},\ref{fLiHfinal},\ref{fLi2finalhyb}, respectively.
As seen from these figures 
the overall agreement between the exact curves (solid lines) and QMC results (open squares) is very good. 
To be more quantitative we have extracted from these curves an estimate of the spectroscopic 
constants $R_e$ (equilibrium distance) and $\omega_e$ (harmonic frequency). To do that, the data
have been fitted with a functional form given by the derivative of a Morse potential curve 
$E(R) = D [\exp{-2 \beta (R-R_e)} - 2 \exp{-2 \beta (R-R_e)}]$ over some interval of distances around 
the equilibrium geometry (R between 1.1 and 2. for H$_2$, between 2.6 and 4. for LiH, and between 
4. and 6. for Li$_2$).
Parameters $D,\beta$, and $R_e$ have been determined via a generalized least-squares fit.
Our results at the VMC, DMC, and Hybrid levels are presented in Table \ref{fitdata} and 
compared to experimental values.\cite{herz} 
As seen from the Table results for the equilibrium distances are excellent. The largest
systematic errors are obtained at the VMC level (relative errors of 4.3$\%$,3.3$\%$, and 5.7$\%$ for H$_2$,LiH, and 
Li$_2$, respectively). A reduction of a factor of about two is gained when DMC calculatons are 
performed. Finally, using the hybrid formula, the exact equilibrium distances are recovered 
within statistical errors (the relative statistical errors being 1.1$\%$,0.3$\%$, and 0.5$\%$ for H$_2$,LiH, and
Li$_2$, respectively). In contrast, results for the harmonic frequencies are less accurate but still satisfactory. 
For the H$_2$ molecule, the exact experimental result is almost recovered within statistical error at the VMC, 
DMC, and Hybrid levels, the relative statistical error being between 3 and 4$\%$. For LiH and Li$_2$ the 
relative statistical errors are of the same order of magnitude. However, a non-negligible systematic 
error of about $10\%$ is found for these molecules. This result illustrates that obtaining accurate 
harmonic frequencies is more difficult than obtaining
accurate equilibrium geometries.

Now, we would like to present a more quantitative discussion 
of the performance of the various force estimators introduced in this work. 
This will allow us to summarize the various aspects of the method,
to present some comparisons with the recent results obtained from the improved estimator 
implicitly used by Casalegno and collaborators\cite{cmr}, and, also, to emphasize 
on some important {\it quantitative} issues. In Table \ref{tableLi2} the systematic and statistical errors 
associated with the various force estimators at the VMC, DMC and Hybrid levels of calculation are presented.
The results shown are for the Li$_2$ molecule at the equilibrium 
geometry (R=5.051 a.u.) where the exact force average, denoted as
$\langle F \rangle_{ex}$, is equal to zero. To allow direct comparisons between force estimators all averages have been 
computed in a common Monte Carlo calculation (identical MC samples).
To compare with, we also give the systematic and statistical errors on the 
total energy. To give a measure of the fluctuations of each estimator the
corresponding variances at the VMC level, $\sigma^2$(VMC), are reported. To facilitate comparisons between 
data all averages (except for the VMC variances) are given with five significant figures after the 
decimal point and all statistical errors are given on the fifth decimal place (magnitude $10^{-5}$). 
The first estimator presented is the bare estimator, $F$ 
[Eq.(\ref{localforcediatomics})]. As already pointed out, this estimator, which 
has an infinite variance, displays very large statistical fluctuations (between two and three orders 
of magnitude with respect to the improved estimators to follow)
and is, therefore, not at all suitable for practical calculations. The second estimator, 
$\tilde{F}_{ZV} [\psi_T, \tilde{\psi}_{min}]$ [Eq.(\ref{f1})],
introduced in our previous work on forces\cite{zvp2} is the simplest estimator having 
a finite variance. However, as explained above, when using such an estimator no control on the systematic error 
exists. The third estimator presented,
$\tilde{F}_{ZVZB} [\psi_T, \tilde{\psi}_{min}]$, is the simplest estimator 
having the ZVZB property. We can see that the introduction of the contribution associated 
with the ZB property, $2(E_L - \langle E_L\rangle) \frac{\tilde{\psi}_{min}}{\psi_T}$ is efficient in 
reducing the bias (the DMC and hybrid errors are roughly divided by a factor two). 
However, as already discussed, the derivative of the 
trial wave function is not correctly reproduced as a function of the 
interatomic distance and the corresponding force curve is not smooth 
(see, Figures \ref{forcevar},\ref{forcedmc}). To get accurate and well-behaved (as a function of $R$) 
values of the force it is important to introduce an auxiliary function close to the 
exact derivative of the wave function. The most simple estimator based on this idea and having a finite 
variance can be constructed by using the minimal form $\tilde{\psi}_{min}$ [Eqs.(\ref{psi}),(\ref{q})]
for the Zero-Variance part and $\tilde{\psi}_{Deriv}$,[Eq.(\ref{psideriv})], for the Zero-Bias part. Such an 
estimator is written as
\begin{equation}
\tilde{F}[\psi_T, \tilde{\psi}_{min} | \tilde{\psi}_{Deriv}]  \equiv F+ \frac{(H-E_L) \tilde{\psi}_{min}}{\psi_T}
+ 2(E_L - \langle E_L\rangle) \frac{\tilde{\psi}_{Deriv}}{\psi_T}
\label{minder}
\end{equation}
Written with our notations this is in fact the estimator implicitly used by Casalegno and collaborators in their 
very recent work\cite{cmr}. In contrast with the other estimators presented here this ``mixed'' estimator 
(different trial functions $\tilde{\psi}$ are used for the ZV and ZB parts) 
has no ZVZB property. Accordingly, our general optimization procedure based on the minimization 
of the improved-estimator variance is no longer meaningful here. Nevertheless, as pointed out 
by Casalegno {\it et al.}, to optimize estimator (\ref{minder}) 
we still have the possibility of optimizing the parameters of the trial wave 
function $\psi_T$ via energy minimization. Such a procedure is justified
because fully-optimized trial wave functions are known to verify the Hellmann-Feynman theorem. 
Statistical errors associated with this estimator are reasonable and roughly similar to those obtained 
with the two previous 
estimators. Systematic errors are also comparable. However, in sharp contrast with all ZVZB-improved-estimators 
introduced in this work, the DMC (and hybrid) calculations do not improve the results and, as seen in the Table 
the hybrid results can be even bad, despite the fact that VMC results are reasonable. 
Regarding the dependence of the results on the interatomic distance 
we have been able to recover a relatively smooth force curve for the smallest molecules H$_2$ and LiH but not for Li$_2$. 
In this latter case, the systematic error is found to be too much sensitive on 
the quality of the optimization of the trial wave function to lead to reliable results. 
The last improved estimator presented in Table \ref{tableLi2} is our best proposal for the 
force estimator, Eq.(\ref{fpsir}). 
We report results with ($\vec{v} \neq 0$) and without ($\vec{v} = 0$) to enlighten the role 
of the coordinate-transformation term. 
As seen, the introduction of the $\vec{v}$-term is extremely efficient in reducing the statistical error. 
For example, at the VMC level the statistical error without this term is 218.10$^{-5}$, while the optimized 
improved estimator using the $\vec{v}$-term is decreased down to 9.10$^{-5}$. 
The reduction gained in statistical error is more than one order of magnitude. 
This remarkable result is general : It is valid for all molecules and all distances treated here. 
Another most important point is that our best improved estimator (\ref{fpsir}) is the only estimator presented in this work 
whose statistical error (here, 9.10$^{-5}$) is (much) smaller than the energy one (here, 32.10$^{-5}$). 
Note that it is also true for the systematic error (whatever the level of calculation). 
Such a result is particularly important since it has been found that a precise control of the magnitude of 
the systematic error through variance minimization 
of the improved force estimators is possible only when such a condition is verified. 
In contrast, when the statistical error on 
the force is larger than the energy error, the variance minimization can lead to various results and to get 
a smooth force curve is very difficult.
Actually, we would like to emphasize that obtaining results of the quality presented in Figure (\ref{fLi2finalhyb})
for the Li$_2$ molecule has only been possible with the improved estimator (\ref{fpsir}). Using other estimators 
we have not been capable of constructing a reasonable force curve (smooth and accurate) for this molecule.

Finally, let us say a word about
the dependence of our results on the optimization process (determination of the optimal parameters entering $\tilde{\psi}$, 
$\psi_T$, and $\vec{v}$ by minimization of the variance of the improved estimator). Clearly, the method presented in this work 
is useful only if the results obtained do not depend too much on the way the optimization is performed and on which particular 
minimum has been found for the variance (as already emphasized, when a large number of parameters are 
considered the location of such a minimum can depend very crucially on the initial conditions 
and/or on the random numbers series used).
To quantify this aspect we have made 9 independent optimizations over 9 independent sets of 2000 walkers for the Li$_2$ molecule 
at the equilibrium geometry with our best estimator. Results show that the VMC average force results may vary in a 
significant way for the different sets of optimized parameters found. In this case, the domain of variation is about twenty 
times the magnitude of the statistical error and about 30$\%$ of the average itself.
However, it has been observed that the DMC and Hybrid averages are much less sensitive on the optimized parameters.
The error on the DMC and hybrid average forces due to an incomplete optimization has been found to be of the order 
of magnitude of the statistical error which is rather small. 

\section{Summary and conclusions}

In this work we have shown how to construct improved VMC or DMC estimators 
for observables. By improved it is meant that, compared to the standard 
bare estimator $O$, the new estimators $\tilde{O}$ have a lower variance 
and a reduced systematic error when averaged over the approximate VMC or 
(fixed-node) DMC probability densities.

At the Variational Monte Carlo level the most general form we propose
for $\tilde{O}$ is given by
\begin{equation}
\tilde{O}_{VMC}[\psi_T,\tilde{\psi},\vec{v}] \equiv  O 
+ \frac{(H-{E_L})\tilde{\psi}}{{\psi_T}}
+2 ({E_L}- \langle E_L\rangle) \frac{ \tilde{\psi}}{\psi_T} +
\frac{\vec{\nabla} [(E_L - \langle E_L\rangle) \psi_T^2 \vec{v}]}
{\psi_T^2}
\label{tildeOconcVMC}
\end{equation}
where averages are defined over the VMC distribution. At the Diffusion Monte Carlo level
the expression proposed is essentially similar, except that 
the average of the local energy entering the definition of $\tilde{O}_{DMC}$ is
defined over the DMC distribution, Eq.(\ref{tildeobias7}).
The various terms defining the improved observables have a well-defined 
physical origin: The three first contributions result from the change 
of the energy average when the magnitude of the observable considered 
as an external field is varied, while the 
last contribution comes from the use of a coordinate transformation 
correlating electron displacements and change of the external field.
The functions $\psi_T$ and $\tilde{\psi}$ appearing in 
the improved observables play the role of trial functions:
$\psi_T$ is the ordinary trial wave function for the ground-state of $H$ and $\tilde{\psi}$ is a 
guess for the derivative of the exact ground-state wave function, 
$\psi_0(\lambda)$, of the perturbed Hamiltonian 
$$
H(\lambda) \equiv H + \lambda O
$$
with respect to $\lambda$ at $\lambda=0$.
When the trial functions are exact, $(\psi_T,\tilde{\psi})=
(\psi_0,\frac{d\psi_0(\lambda)}{d\lambda} 
\big|_{\lambda =0})$, the improved estimator reduces to a constant, namely 
the exact average for the observable. In that case both statistical 
and systematic errors vanish. We have called this 
remarkable property ``Zero-Variance Zero-Bias property''(ZVZB).
In the neighborhood of the exact solution, 
a local expansion of the various quantities obtained from the 
approximate guess $(\psi_T,\tilde{\psi})$ can be 
done. It is found that there is a {\it quadratic} 
behavior in the errors $\delta \psi= \psi_T-\psi_0$ and 
$\delta \psi^\prime= \tilde{\psi}-\psi_0^\prime$. At the VMC level it 
reads
\begin{equation}
\sigma^2(\tilde{O}) \equiv \langle (\tilde{O} - \langle \tilde{O} \rangle_{{\psi_T}^2})^2 
\rangle_{{\psi_T}^2} \sim O[\delta \psi \delta \psi^\prime]
\label{conc1}
\end{equation}
and
\begin{equation}
\Delta_{\tilde O} \equiv
\langle \tilde O \rangle_{{{\psi_T}}^2} - \langle O \rangle_{\psi_0^2}
\sim O[\delta \psi \delta \psi^\prime],
\label{conc2}
\end{equation}
with a similar result in the DMC case.
This important result generalizes the well-known quadratic Zero-Variance 
Zero-Bias property of the energy where the local energy, 
$E_L = H\psi_T/\psi_T$ plays the role of the improved estimator:
\begin{equation}
\sigma^2(E_L) \sim O[{\delta \psi}^2]
\label{conc3}
\end{equation}
and
\begin{equation}
\Delta_{E_L} \sim O[{\delta \psi}^2].
\label{conc4}
\end{equation}
In the case of the energy we can write a Zero-Variance Zero-Bias equation 
defining the optimal trial wave function by imposing that the local energy 
reduces to a constant, namely the
exact energy
\begin{equation}
\frac{H\psi_T}{\psi_T} = E_0.
\label{zvzbeqe0}
\end{equation}
Of course, this equation is nothing but the Schroedinger equation. 
Here, the Zero-Variance Zero-Bias equation for the observable is obtained 
by imposing that the improved observable $\tilde{O}$ reduces to the 
exact average
\begin{equation}
\tilde{O}= \langle O \rangle_{\psi_0^2}.
\end{equation}
By optimizing the three quantities $(\psi_T,\tilde{\psi},\vec{v})$ so that
fluctuations of $\tilde{O}$ are minimal we can obtain the optimal improved estimator
for the observable. In practice, it is done in two steps. First, functional forms 
for the trial functions $\psi_T$ and $\tilde{\psi}$ are chosen in order to reproduce 
the best as possible the exact solution of the zero-variance equations. The choice 
of the vector field $\vec{v}$ is done on physical grounds: It corresponds to the 
electron-coordinate transformation, $\vec{y}=\vec{x}+\lambda \vec{v}(\vec{x})
+O(\lambda^2)$, correlating as much as possible the electron displacements and 
the change of density associated with the external field defined by the bare 
observable. Second, the various parameters entering the three quantities 
are optimized by minimizing the fluctuations of $\tilde{O}$ over a large but 
finite number of configurations (typically, several thousands) drawn according to 
the VMC or DMC distributions.

It is important to emphasize that by using the improved estimators presented here 
it is possible to get an accuracy on expectation values of observables which is
comparable to the very good one obtained for total energies. As it can be seen from 
Eqs.(\ref{conc1},\ref{conc2},\ref{conc3},\ref{conc4}) this is true when we are able 
to reduce the error on the derivative of the wave function at the level 
of the error on the wave function itself, that is $\delta \psi \sim {\delta \psi}^\prime$.

Another fact worth pointing out is that there is not an unique way of constructing 
improved estimators. Here, we have built our estimators by considering the derivative 
of the variational, Eq.(\ref{evprime}), or the exact DMC energy average, Eq.(\ref{e0vprime}), 
We have also considered the possibility of making a coordinate 
transformation before making the derivative, Eqs.(\ref{eJacprime},\ref{evprime2}). 
Of course, we can think of many other choices and/or transformations. Ultimately, 
the better strategy will depend very much on the specific problem considered.

Finally, in order to go beyond VMC or DMC calculations, 
we have shown that the reduction of error of one order in $\delta \psi$ associated 
with the popular ``hybrid''(or ``second-order'') formula mixing DMC 
and VMC averages can be generalized to the case of our improved estimators:
\begin{equation}
2 \langle \tilde{O}_{DMC} \rangle_{\psi_T  \psi_0} -
  \langle \tilde{O}_{VMC} \rangle_{\psi_T^2} \sim \langle \psi_0 | O |\psi_0 \rangle.
\label{hybridconc}
\end{equation}

As an important application we have applied our formalism to the case of the 
computation of forces for some simple diatomic molecules. In our preceding work 
on forces\cite{zvp2} we have focused our attention only on the zero-variance part of 
the problem. More precisely, we have employed i) a simplified version of the renormalized force, Eq.(\ref{f1}) 
ii.) the minimal expression for $\tilde{\psi}$ leading to a finite variance, 
Eqs.(\ref{psi},\ref{q}), and iii.) the hybrid formula mixing 
VMC and DMC calculations, Eq.(\ref{hybridconc}). Results obtained for 
the vanishing force at the equilibrium distance for a number of 
small diatomic molecules were reasonably good.
Here, we have illustrated that such a strategy is in fact not valid for 
describing the global shape of the force curve. It has been shown that results 
depend very much on the trial wave function used and, particularly, on the quality 
of the optimization process of the numerous parameters of the trial wave function. 
As a result, the force curves obtained are not regular as a function of the 
interatomic distance and important spectroscopic quantities such as the equilibrium 
distance $R_e$ and the harmonic frequency $\omega_e$
cannot be obtained reliably. To get accurate curves
we need not only to have a small amount of statistical fluctuations but also 
a control of the systematic error.
By exploiting the general ZVZB principle presented in this work it has been shown that 
obtaining accurate curves is now possible. The basic ingredients are: i.) the use of 
a trial wave function for the derivative, $\tilde{\psi}$, built as a 
finite-difference of the trial wave function with respect to the nuclear coordinate ii.) the use of a
coordinate transformation in the spirit of the work of Filippi and Umrigar\cite{fil3} and, finally, iii.) 
the systematic minimization of the variance of the improved estimator 
with respect to all the parameters entering the two trial functions ($\psi_T,\tilde{\psi}$), and the vector field, $\vec{v}$,
associated with the coordinate transformation. Let us emphasize that to get a well-balanced optimization of the 
two trial functions (leading to smooth curves for the forces) , it is essential to reduce the variance of the improved 
estimator for the force at the level of the variance of the local energy. Although such a condition may appear as 
very difficult to fulfill (local energies have usually very small variances), we have shown that
it is in fact possible thanks to the coordinate-transformation term.
Such a result is remarkable and is certainly one of most important 
practical aspect of the approach proposed in this work.

Finally, let us remark that the price to pay with respect to the minimal scheme presented in our previous work\cite{zvp2}
lies on the need of computing about $3N_{nucl}$ local energies to calculate the various
components of the force. However, we do not think it represents a major difficulty for 
the realistic applications to come. Indeed, the few-percent accuracy needed on average forces
will be obtained with relatively small statistics and, therefore, 
it will not be necessary to compute the force vector at each Monte Carlo step
(the expensive $3N_{nucl}$-local energy-calculation step will be done rarely). 
In addition to this, in applications where the nuclear geometry is varied 
during the simulation (Molecular Dynamics-type applications) it should also be possible to use 
suitable re-actualization schemes to avoid re-computing entirely the $3N_{nucl}$ local energies 
for close nuclear configurations. Of course, the validity of these various strategies as well as
the quality of the improved estimators presented in this work need now to be 
checked for realistic applications involving many nuclear degrees of freedom.

{\bf Acknowledgments}
This work was supported by the ``Centre National de la Recherche 
Scientifique'' (CNRS).\\

\textwidth=18truecm
\begin{table}[htp]
\caption{VMC, DMC, and Hybrid estimates of the equilibrium geometry $R_e$ (a.u.)
and harmonic frequency $\omega_e$ (cm$^{-1}$). The atomic isotopic masses taken$^a$ are 1.007825035 
amu for $^1$H and 7.0160030 amu for $^7$Li.}
\begin{tabular}{llll}
\multicolumn{1}{c}{               } &
\multicolumn{1}{c}{H$_2$          } &
\multicolumn{1}{c}{LiH            } &
\multicolumn{1}{c}{Li$_2$         } \\
\hline
R$_e$ (VMC) & 1.463(12)    & 3.111(17)   &  5.346(27) \\
R$_e$ (DMC) & 1.426(13)    & 3.056(6)   &   5.200(16) \\
R$_e$ (Hybrid) & 1.395(15)    & 3.001(15)   &  5.068(27)  \\
R$_e$ (Exp.)$^b$ & 1.401    & 3.015   &  5.051             \\
\hline
$\omega_e$ (VMC) & 4194(130) & 1559(40)   &    366(9)           \\
$\omega_e$ (DMC) &  4432(165) & 1549(22)   &   373(5)            \\
$\omega_e$ (Hybrid) & 4662(205) & 1519(31)   &  387(8)         \\
$\omega_e$ (Exp.)$^b$ & 4395.2     & 1405.65   &  351.4            \\
\end{tabular}
\raggedright
\small{}
$^a$ Ref.\cite{amu}\\
$^b$ Ref.\cite{herz}\\
\label{fitdata}
\end{table}

\begin{table}[htp]
\caption{VMC, DMC, and Hybrid systematic (bias) and statistical errors for the total
energy and various force estimators for Li$_2$ at R=5.051 a.u. The VMC variances,
$\sigma^2$ (VMC), are also given. $\langle E_L\rangle_{ex}$ and $\langle F\rangle_{ex}$ denote the exact
total energy, $\langle E_L\rangle_{ex}$=-14.9954 a.u. and exact force $\langle F\rangle_{ex}=0.$
(equilibrium geometry), respectively. To facilitate comparisons between
energy and force results all averages are given with five significant figures after the
decimal point and statistical errors are given on the fifth decimal place (magnitude $10^{-5}$).
Statistical errors on VMC variances are on the last digit.}
\begin{tabular}{ccccc}
\multicolumn{1}{c}{Estimator      } &
\multicolumn{1}{c}{VMC Average    } &
\multicolumn{1}{c}{DMC Average    } &
\multicolumn{1}{c}{Hybrid         } &
\multicolumn{1}{c}{$\sigma^2$(VMC)} \\
\hline
$E_L - \langle E_L\rangle_{ex}$
& 0.03871(32)    & 0.00531(50)   & -               & 0.113(5)    \\
$^a F - \langle F \rangle_{ex} $
&0.18217(23216) & 0.15462(12293)& 0.12707(33185) & $+\infty$    \\
$^b \tilde{F}_{ZV} [\psi_T, \tilde{\psi}_{min}] - \langle F \rangle_{ex}$
& -0.06352(84)& -0.04003(151)&-0.01654(313)& 1.27(5) \\
$^c \tilde{F}_{ZVZB} [\psi_T, \tilde{\psi}_{min}]- \langle F\rangle_{ex}$
&  -0.05802(104) &-0.02484(184)&0.00834(382) & 1.3(2)      \\
$^d \tilde{F}[\psi_T,\tilde{\psi}_{min} | \tilde{\psi}_{Deriv}]- \langle F\rangle_{ex}$
& 0.00619(109) &0.02993(187) &0.05367(390) & 2.8(1)      \\
$^e \tilde{F}_{ZVZB} [\psi_T, \tilde{\psi}_{Deriv},\vec{v}=0]- \langle F \rangle_{ex}$
& 0.00871(218)&0.00474(147)&0.00077(366) & 14(3)       \\
$^e \tilde{F}_{ZVZB} [\psi_T,\tilde{\psi}_{Deriv},\vec{v}]- \langle F\rangle_{ex}$
&0.00692(9) &0.00358(19)&0.00024(39)& 0.016(1)    \\
\end{tabular}
\raggedright
a. Eq.(\ref{localforcediatomics})\\
b. Eq.(\ref{f1})\\
c. Eq.(\ref{f2})\\
d. Eq.(\ref{minder})\\
e. Eq.(\ref{fpsir})\\
\label{tableLi2}
\end{table}
\textwidth=16truecm

\newpage
\begin{center}
FIGURE CAPTIONS
\end{center}

\begin{center}
\begin{figure}[htp]
\includegraphics[height=18cm,width=18cm,angle=0]{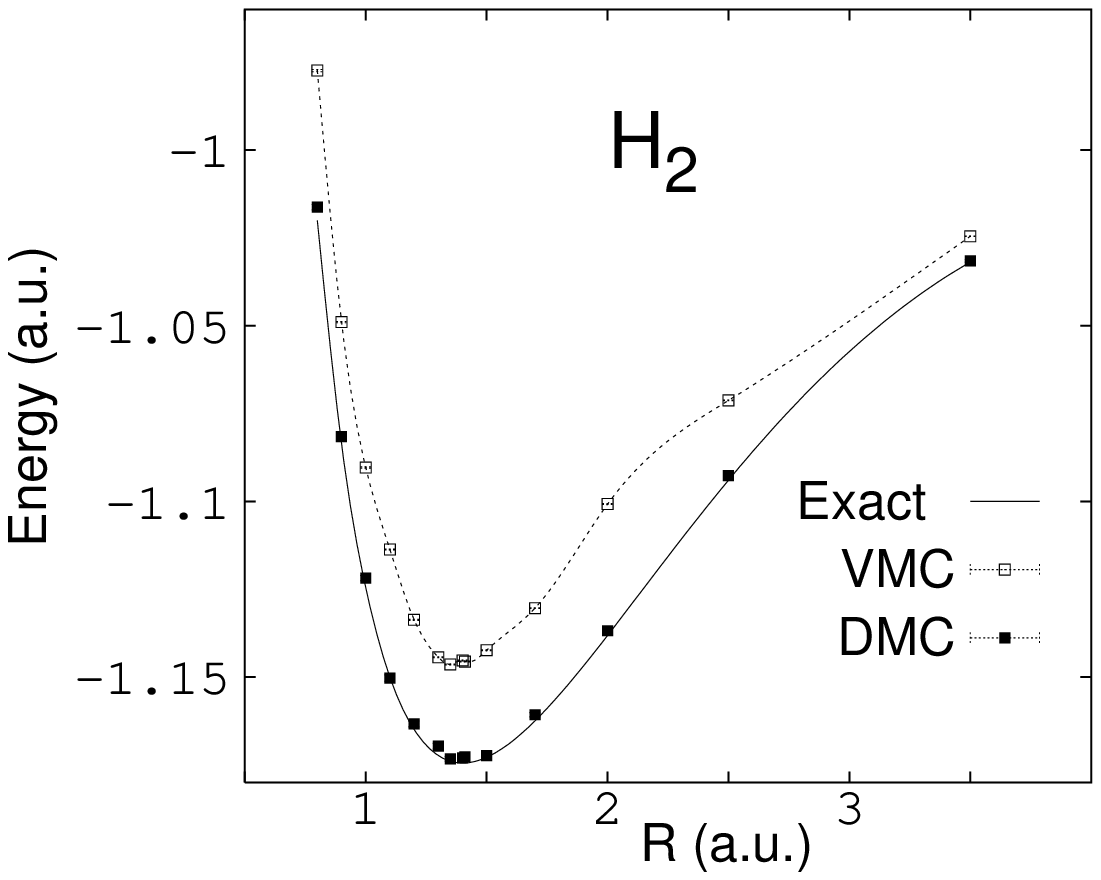}
\caption{}
\label{figureEH2}
\end{figure}
\end{center}

\begin{center}
\begin{figure}[htp]
\includegraphics[height=18cm,width=18cm,angle=0]{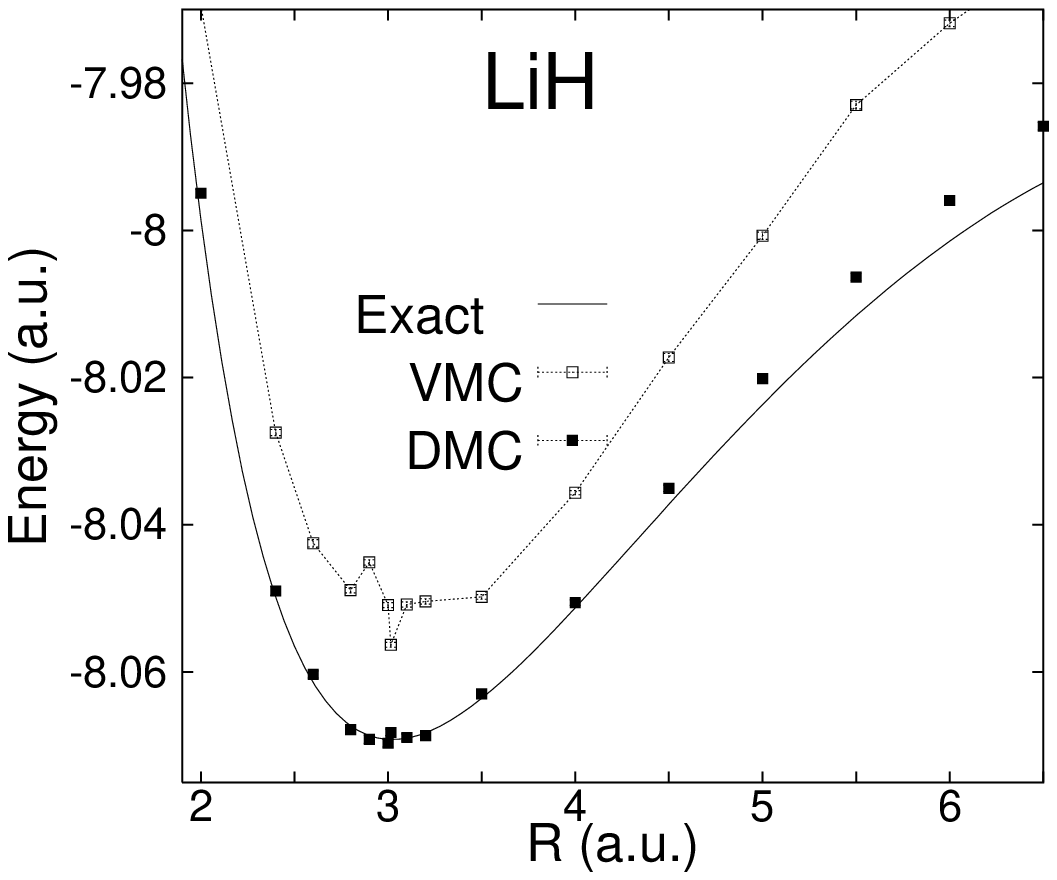}
\caption{}
\label{figureELiH}
\end{figure}
\end{center}

\begin{center}
\begin{figure}[htp]
\includegraphics[height=18cm,width=18cm,angle=0]{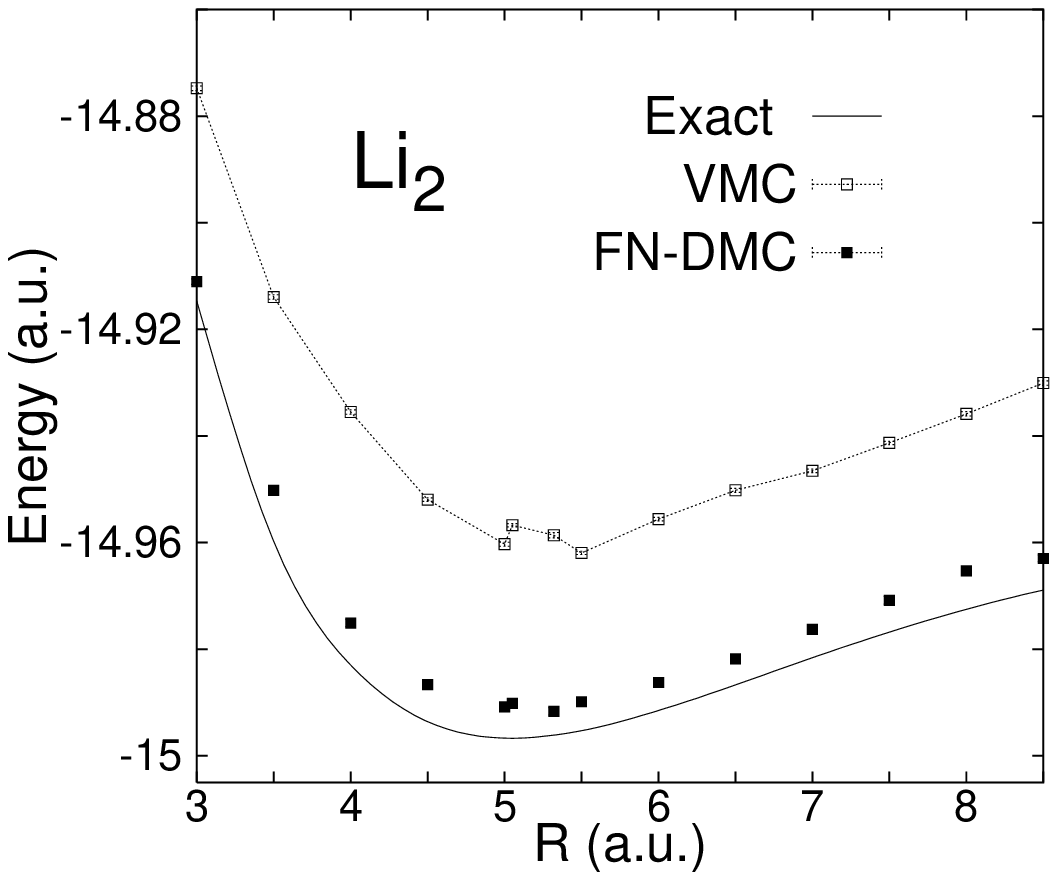}
\caption{}
\label{figureELi2}
\end{figure}
\end{center}

\begin{center}
\begin{figure}[htp]
\includegraphics[height=18cm,width=18cm,angle=0]{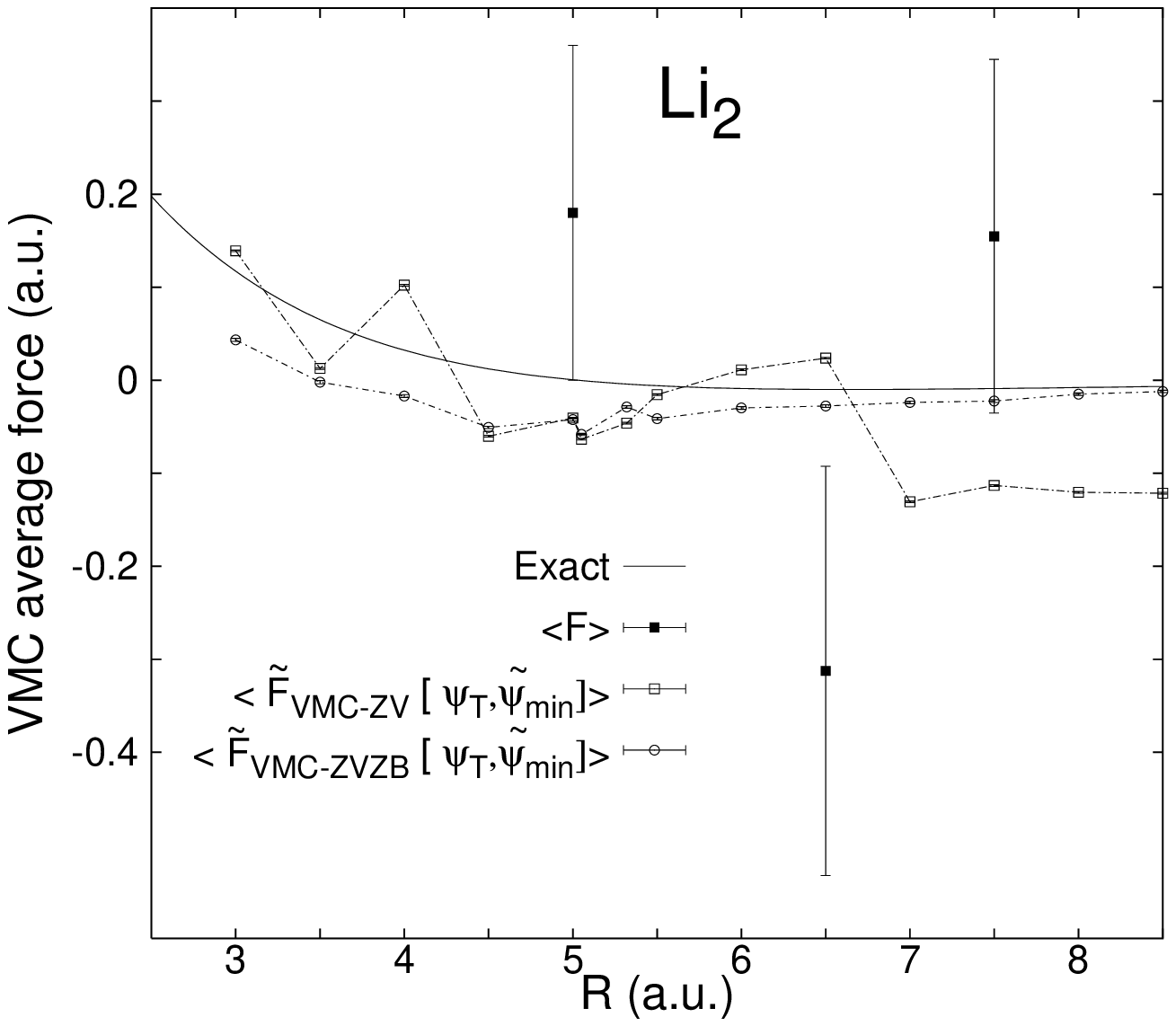}
\caption{}
\label{forcevar}
\end{figure}
\end{center}

\begin{center}
\begin{figure}[htp]
\includegraphics[height=18cm,width=13cm,angle=-90]{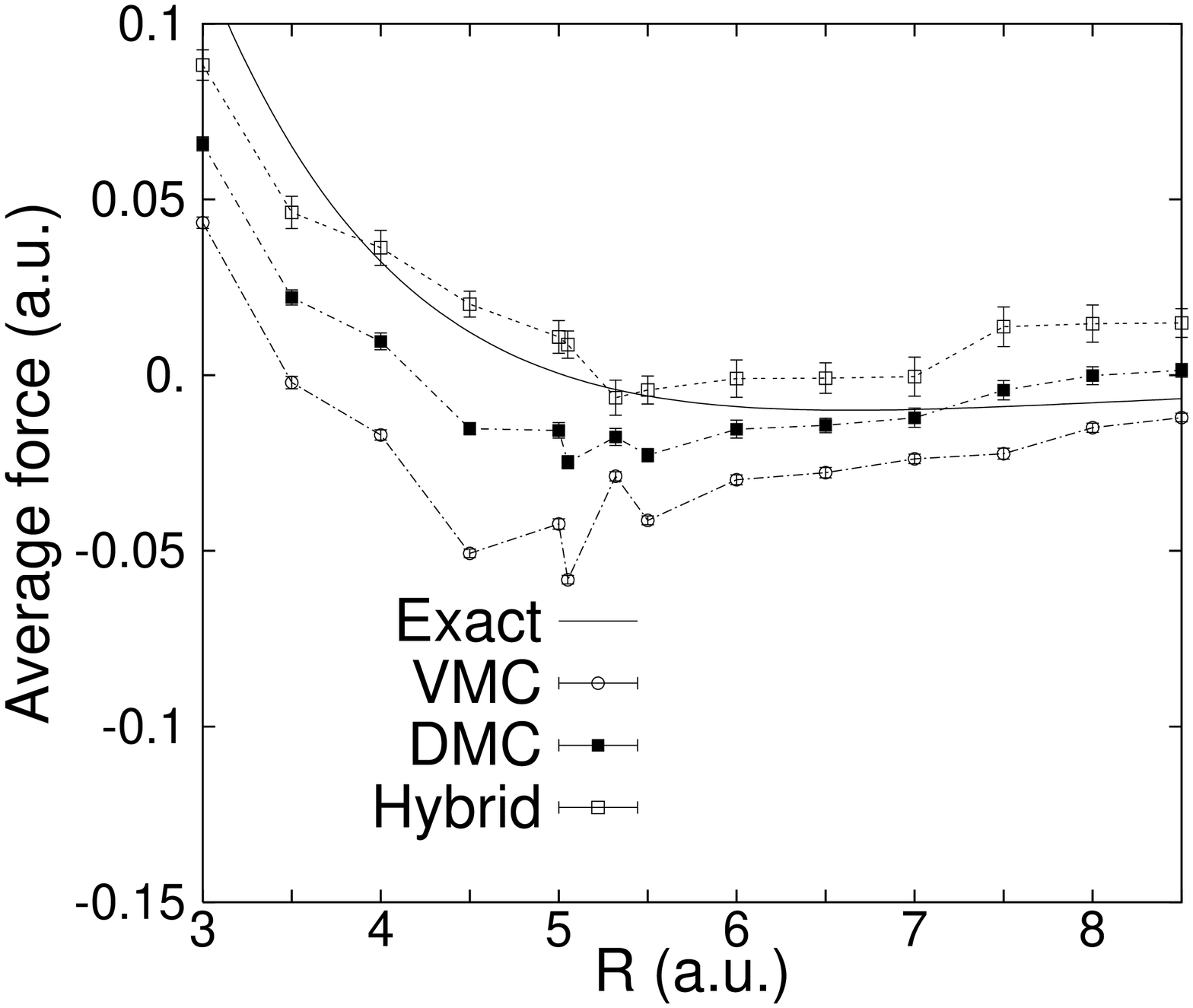}
\caption{}
\label{forcedmc}
\end{figure}
\end{center}

\begin{center}
\begin{figure}[htp]
\includegraphics[height=18cm,width=13cm,angle=-90]{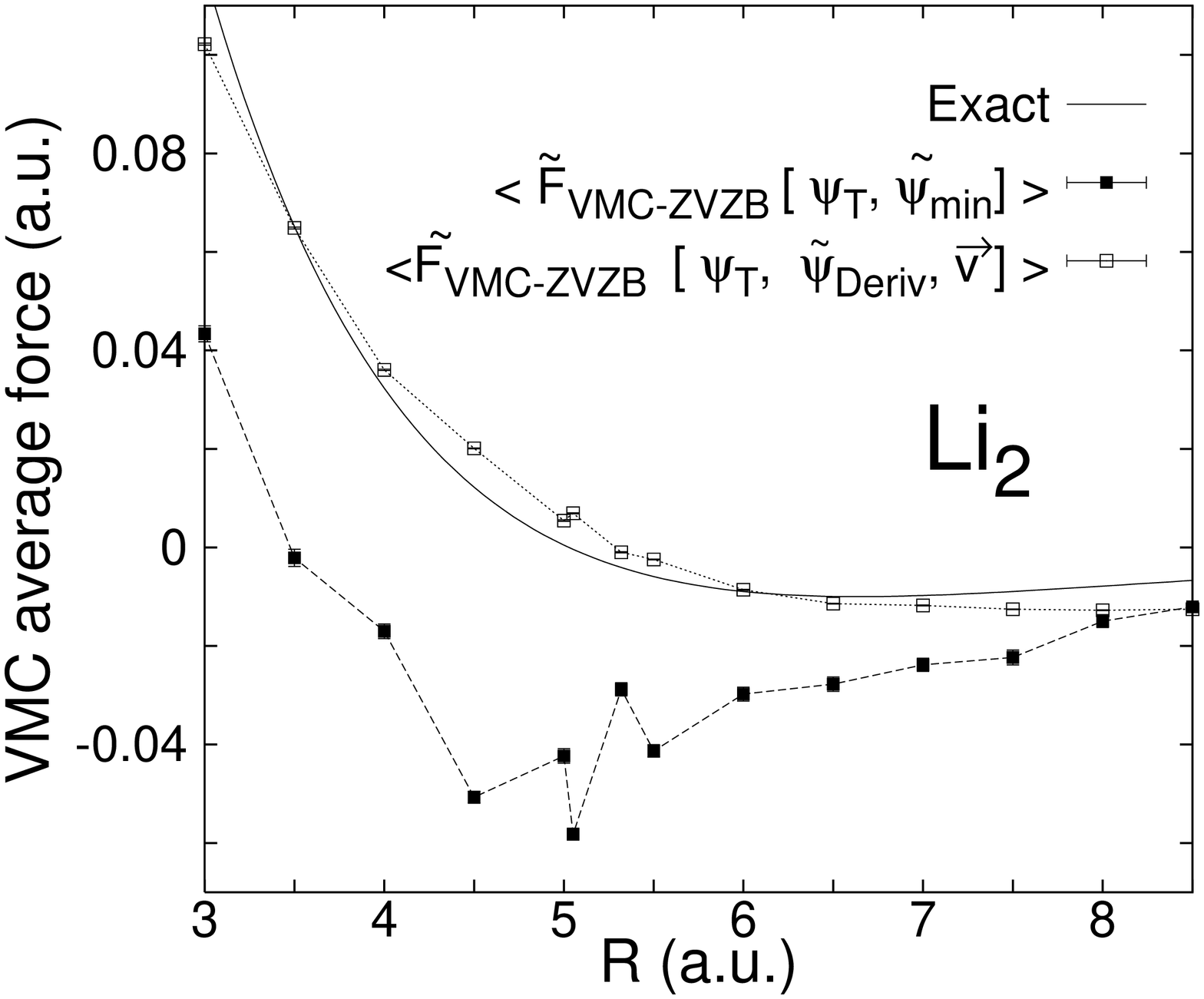}
\caption{}
\label{fLi2psiR}
\end{figure}
\end{center}

\begin{center}
\begin{figure}[htp]
\includegraphics[height=18cm,width=13cm,angle=-90]{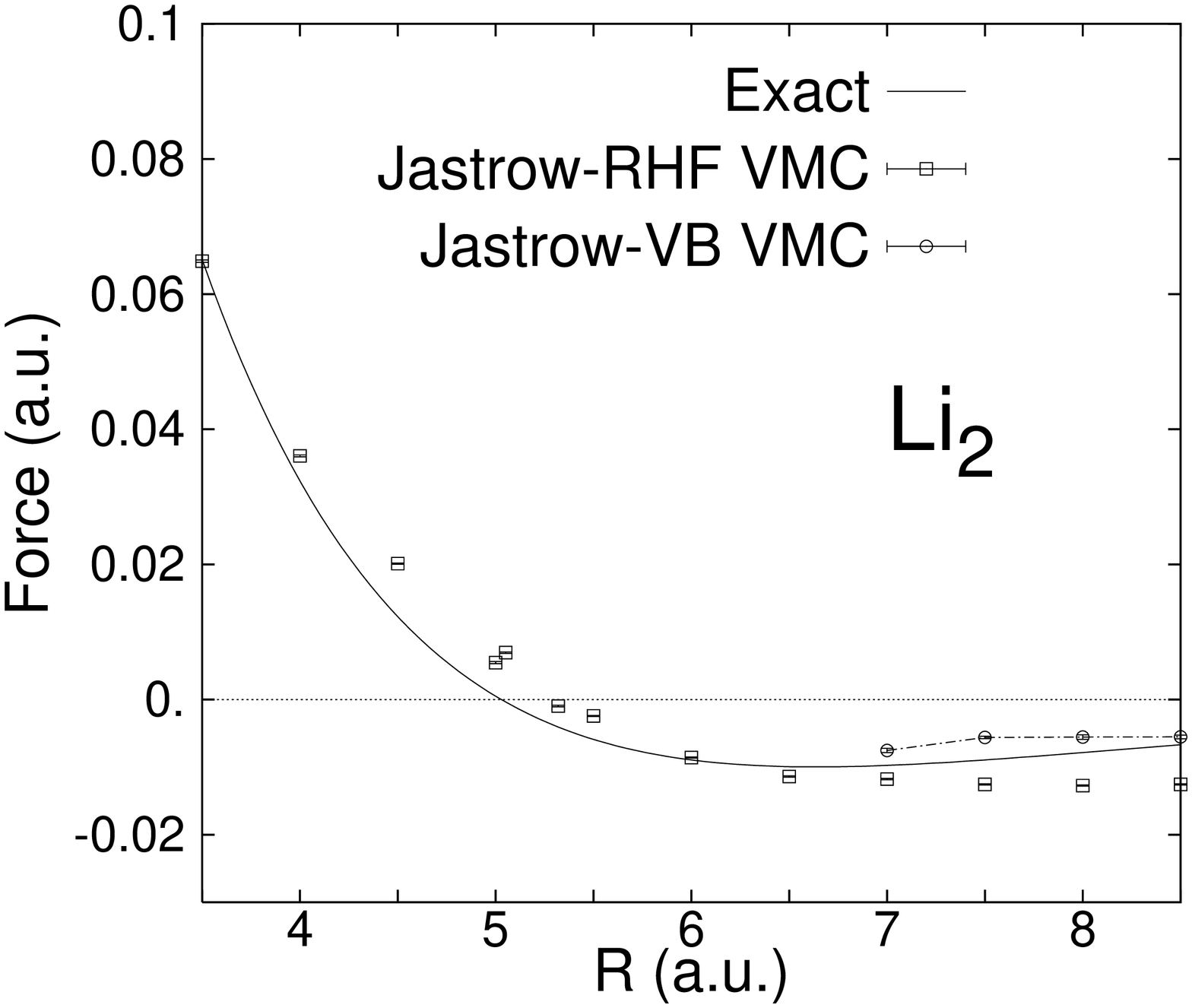}
\caption{}
\label{fLi2finalvmc}
\end{figure}
\end{center}

\begin{center}
\begin{figure}[htp]
\includegraphics[height=18cm,width=13cm,angle=-90]{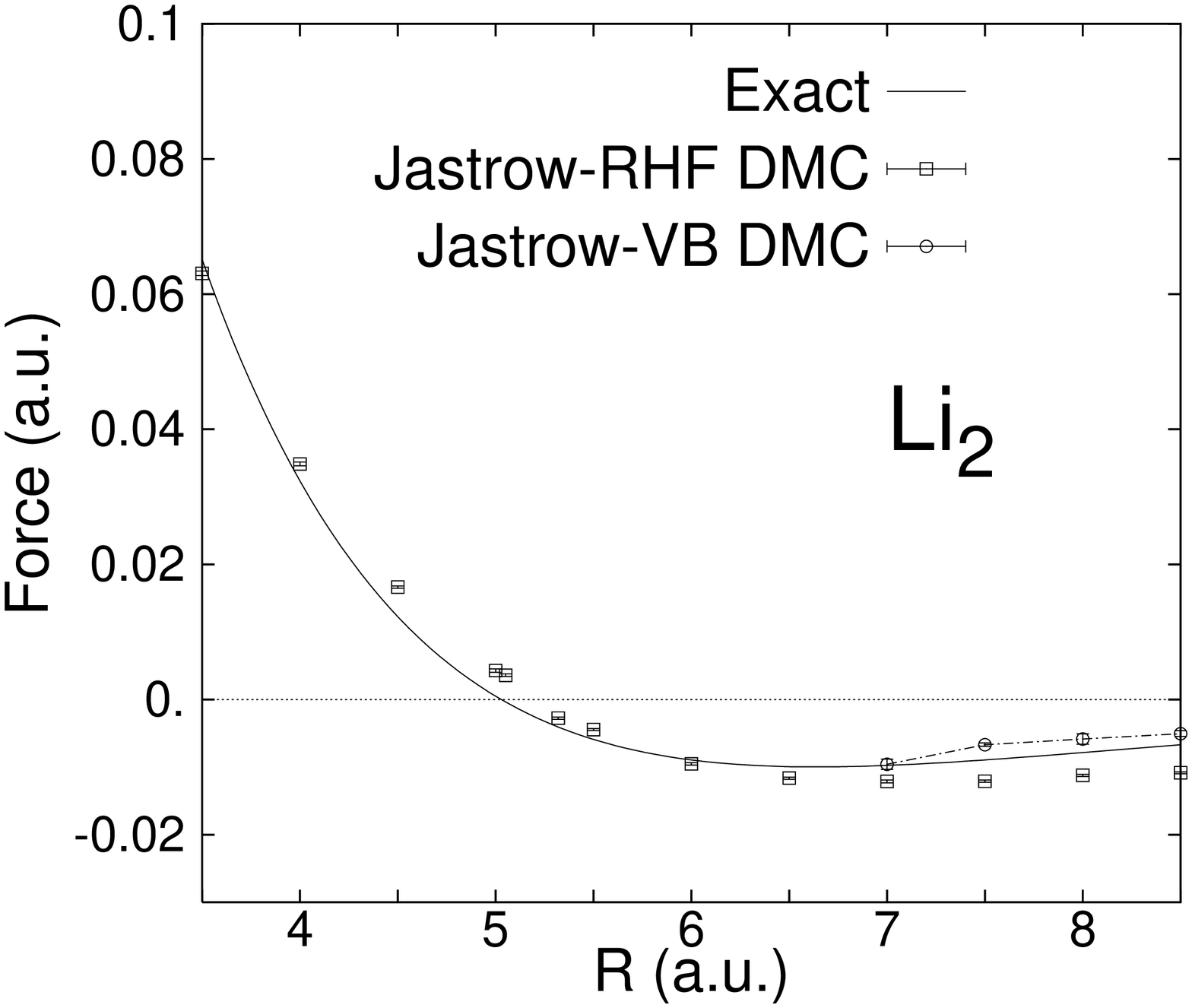}
\caption{}
\label{fLi2finaldmc}
\end{figure}
\end{center}

\begin{center}
\begin{figure}[htp]
\includegraphics[height=18cm,width=13cm,angle=-90]{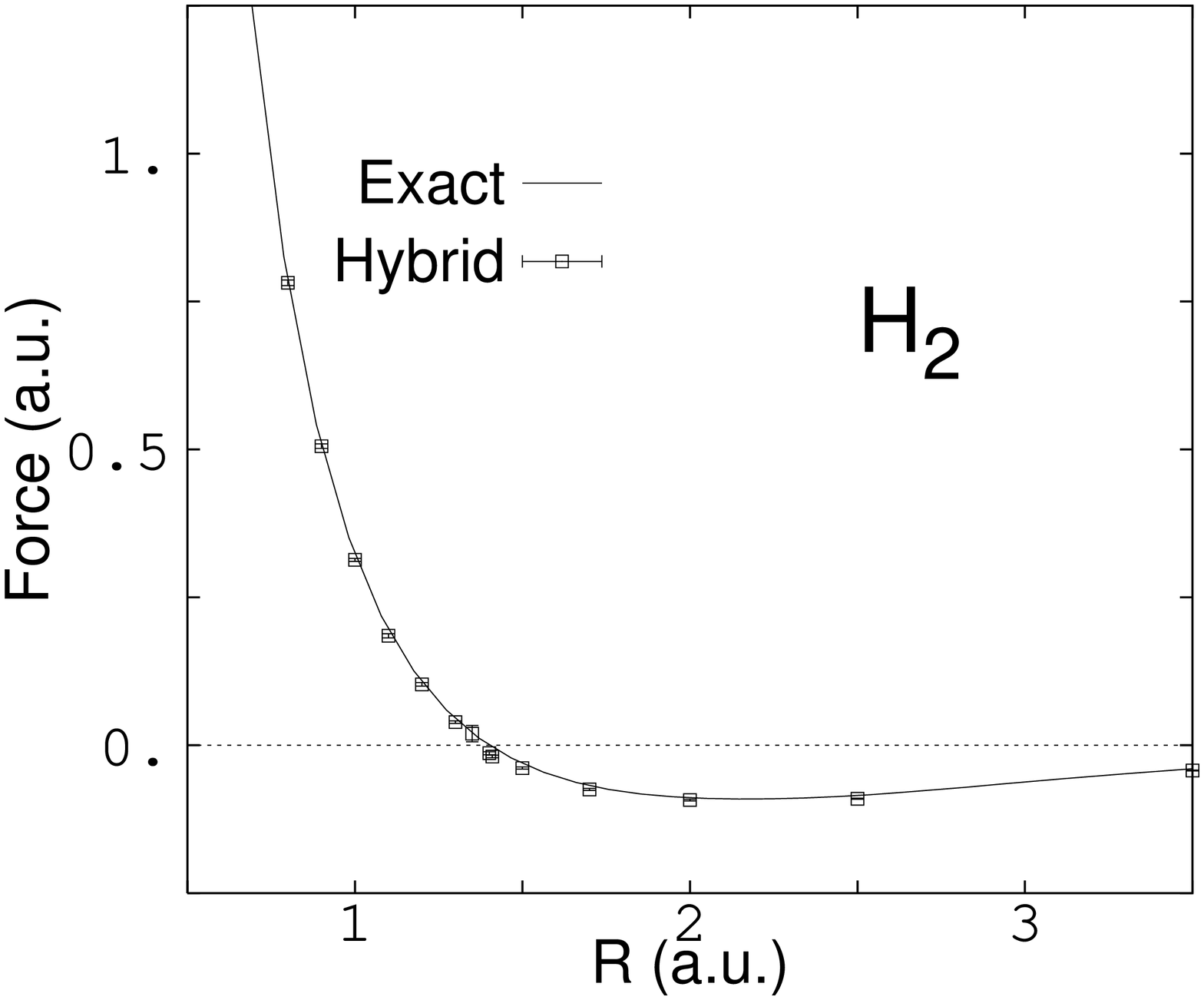}
\caption{}
\label{fH2final}
\end{figure}
\end{center}

\begin{center}
\begin{figure}[htp]
\includegraphics[height=18cm,width=13cm,angle=-90]{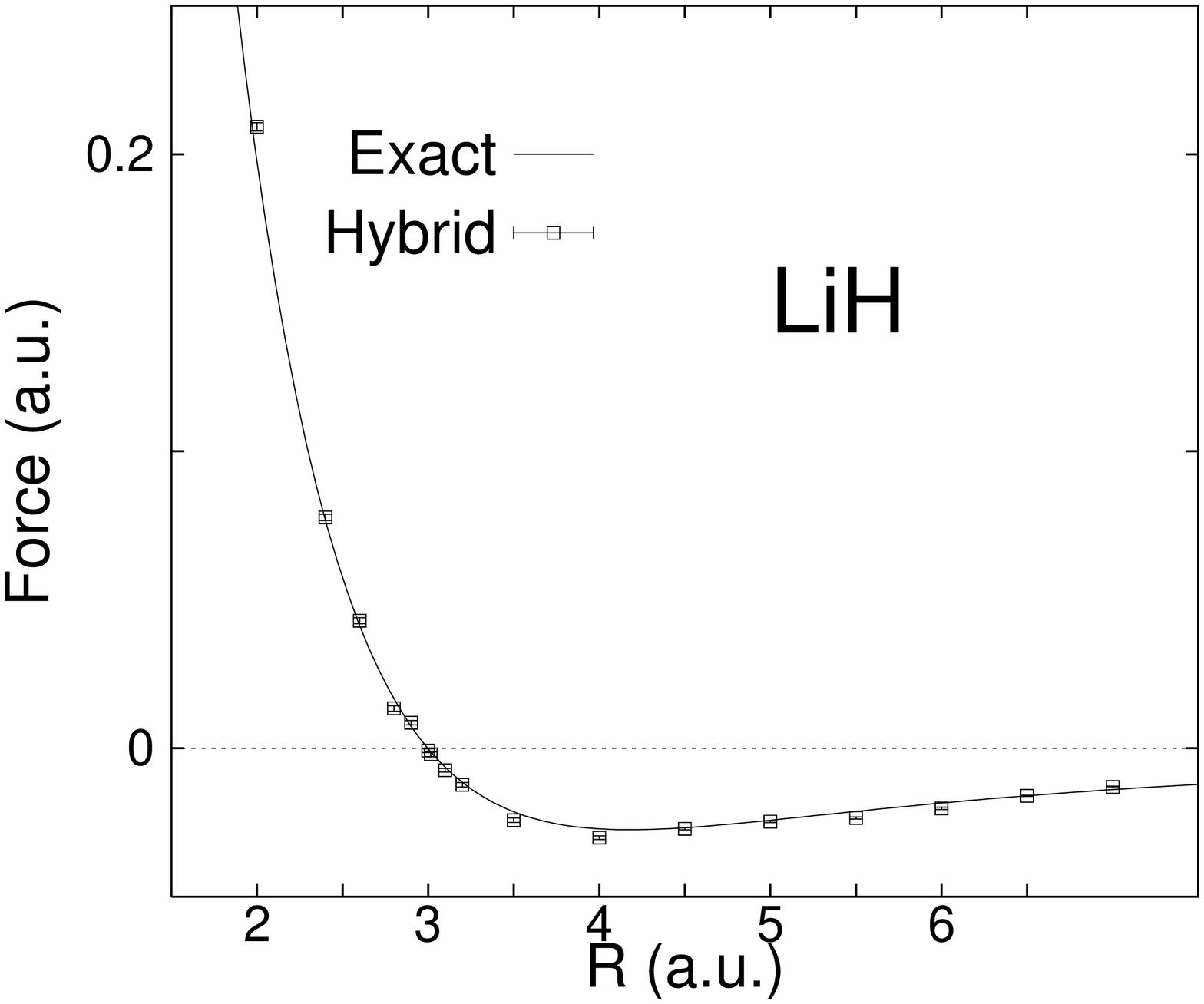}
\caption{}
\label{fLiHfinal}
\end{figure}
\end{center}

\begin{center}
\begin{figure}[htp]
\includegraphics[height=18cm,width=13cm,angle=-90]{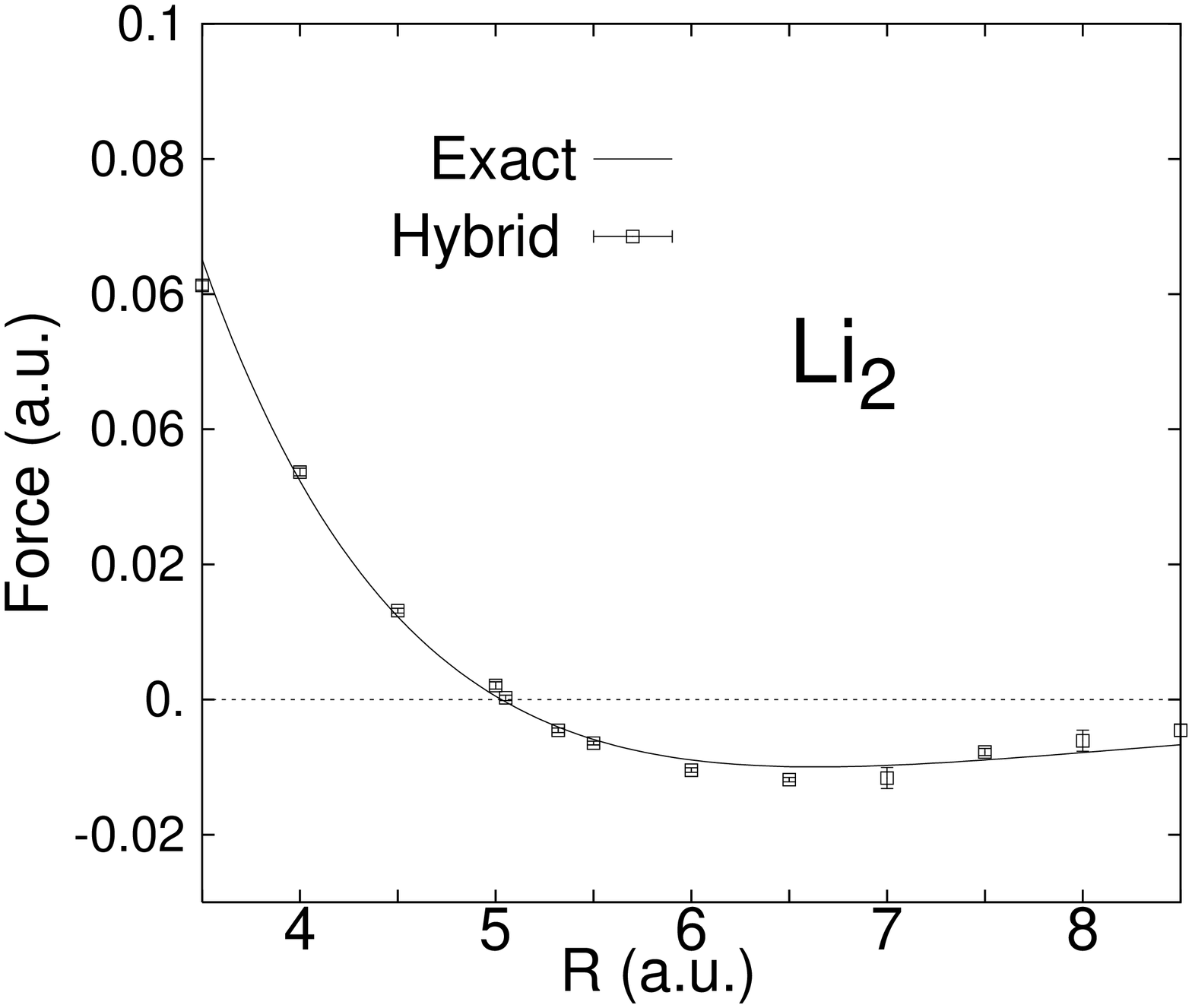}
\caption{}
\label{fLi2finalhyb}
\end{figure}
\end{center}

\begin{itemize}
\item Fig.\ref{figureEH2} H$_2$ molecule. Variational Monte Carlo (VMC) energies (open squares), Diffusion Monte Carlo (DMC)
energies (filled squares) and exact non-relativistic curve (solid line). The dotted line between
VMC results is a simple linear interpolation to guide the eye.

\item Fig.\ref{figureELiH}
LiH molecule. Variational Monte Carlo (VMC) energies (open squares), fixed-node
Diffusion Monte Carlo (DMC) energies (filled squares),
and exact non-relativistic curve (solid line). The dotted line between
VMC results is a simple linear interpolation to guide the eye.

\item Fig.\ref{figureELi2} Li$_2$ molecule. Variational Monte Carlo (VMC) energies (open squares),
fixed-Node Diffusion Monte Carlo (DMC) energies (filled squares),
and exact non-relativistic curve (solid line). The dotted line between
VMC results is a simple linear interpolation to guide the eye.

\item Fig.\ref{forcevar} Various VMC average forces for Li$_2$. Filled squares 
with large error bars: $\langle F \rangle$, Eq.(\ref{localforcediatomics}).
Open squares joined by the dashed line: $\langle \tilde{F}_{VMC-ZV}[\psi_T,\tilde{\psi}_{min}]\rangle$, Eq.(\ref{f1});
Circles joined with the dotted line: $\langle \tilde{F}_{VMC-ZVZB}[\psi_T,\tilde{\psi}_{min}]\rangle$,
Eq.(\ref{f2}). Solid line: exact non-relativistic force curve.

\item Fig.\ref{forcedmc} Li$_2$ molecule. Average forces using $\tilde{F}_{ZVZB} (\tilde{\psi_T},
\tilde{\psi_{min}})$,Eq.(\ref{f2},\ref{f3},\ref{hybridfinal2}).
VMC average: lowest curve with open circles. DMC average: intermediate curve with filled squares.
Hybrid average: highest curve with open squares. Solid line: exact non-relativistic force curve.
Dashed lines between QMC results are a simple linear interpolation to guide the eye.

\item Fig.\ref{fLi2psiR} VMC force for Li$_2$.
Lowest irregular curve with filled squares: $\langle \tilde{F}_{VMC-ZV}[\psi_T,\tilde{\psi}_{min}]\rangle$, Eq.(\ref{f1}).
Upper curve with open squares: $\langle \tilde{F}_{VMC-ZV}[\psi_T,\tilde{\psi}_{Deriv},\vec{v}]\rangle$, Eq.(\ref{fpsir}).
Solid line: exact non-relativistic force curve.

\item Fig.\ref{fLi2finalvmc} VMC force for Li$_2$.
Open squares: Average VMC forces from estimator (\ref{fpsir}) using the Jastrow-RHF one-determinant wave function.
Open circles: Average VMC forces from estimator (\ref{fpsir}) using the Jastrow-VB six-determinant wave function.
Solid line: exact non-relativistic force curve.

\item Fig.\ref{fLi2finaldmc} DMC force for Li$_2$.
Open squares: Average fixed-node DMC forces using the Jastrow-RHF one-determinant wave function.
Open circles: Average fixed-node DMC forces using the Jastrow-VB six-determinant wave function.
Solid line: exact non-relativistic force curve.

\item Fig.\ref{fH2final} Hybrid force for H$_2$.
Solid line: exact non-relativistic force curve.

\item Fig.\ref{fLiHfinal} Hybrid force for LiH.
Solid line: exact non-relativistic force curve. 

\item Fig.\ref{fLi2finalhyb} Hybrid force for Li$_2$.
Solid line: exact non-relativistic force curve.

\end{itemize}

\end{document}